\documentclass[a4paper,11pt]{article}
\usepackage{graphicx,epsfig}
\usepackage{grffile}
\usepackage{amsmath,amssymb}
\usepackage{array}
\usepackage{color}
\usepackage{slashed}
\usepackage{multirow}
\usepackage[normalem]{ulem}
\usepackage{cite}

\title{\Large\bf\boldmath Matter effects in neutrino visible decay at future long-baseline experiments}
\date{}
\author{M.~V.~Ascencio-Sosa, A.~M.~Calatayud-Cadenillas, A.~M.~Gago, J.~Jones-P\'erez
\\[0.5 cm]
{\em\normalsize Secci\'on F\'isica, Departamento de Ciencias, Pontificia Universidad Cat\'olica del Per\'u,} \\
{\em\normalsize  Apartado 1761, Lima, Peru} \\
}

\begin{document}
 
\maketitle 
 
\begin{abstract}
\noindent
Neutrino visible decay in the presence of matter is re-evaluated. We study these effects in two future long-baseline experiments where matter effects are relevant: DUNE (1300~km) and a hypothetical beam aimed towards ANDES (7650~km). We find that matter effects are negligible for the visible component of neutrino decay at DUNE, being much more relevant at ANDES. We perform a detailed simulation of DUNE, considering $\nu_\mu$ disappearance and $\nu_e$ appearance channels, for both FHC and RHC modes. The sensitivity to the decay constant $\alpha_3$ can be as low as $2\times10^{-6}$~eV$^2$ at 90\% C.L., depending on the neutrino masses and type of coupling. We also show the impact of neutrino decay in the determination of $\theta_{23}$ and $\delta_{\rm CP}$, and find that the best-fit value of $\theta_{23}$ can move from a true value at the lower octant towards the higher octant.
\end{abstract} 
 
\section{Introduction} 
 
The neutrino oscillation phenomenon is the only firm evidence of physics beyond the Standard Model. It is then of interest to study at depth if neutrino oscillations could give further information regarding a more complete description of Nature. In this case, one would hope that measurements in neutrino experiments would eventually deviate from the expectations of the standard neutrino oscillation paradigm. Such kind of deviations have been extensively studied in the literature, notable examples being decoherence~\cite{Gago:2000qc,Gago:2002na,Fogli:2003th,Morgan:2004vv,Lisi:2000zt,Hooper:2004xr,Farzan:2008zv,Oliveira:2010zzd,robertoPhd,Oliveira:2013nua,Berryman:2014yoa,Carpio:2017nui,Coloma:2018idr} and non-standard interactions~\cite{GonzalezGarcia:1998hj, Bergmann:2000gp, Guzzo:2004ue, Gago:2001si, Gago:2001xg, Fogli:2007tx, Ohlsson:2012kf, Esmaili:2013fva}. 

Light neutrino decay constitutes a third way of producing changes in neutrino oscillation experiments. In particular, models where the neutrino decay is due to its coupling to a massless scalar, called a Majoron, have been analyzed in detail by many authors~\cite{Berryman:2014yoa, Frieman:1987as, Raghavan:1987uh, Berezhiani:1991vk, Berezhiani:1992ry, Berezhiani:1993iy, Barger:1999bg, Lindner:2001fx, Beacom:2002cb, Joshipura:2002fb, Bandyopadhyay:2002qg, Ando:2004qe, Fogli:2004gy, PalomaresRuiz:2005vf, GonzalezGarcia:2008ru, Maltoni:2008jr, Baerwald:2012kc, Meloni:2006gv, Das:2010sd, Dorame:2013lka, Gomes:2014yua, Berryman:2014qha, Picoreti:2015ika, Abrahao:2015rba, Bustamante:2016ciw, Gago:2017zzy, Coloma:2017zpg, Choubey:2017dyu, Choubey:2017eyg}. In these kind of models, neutrinos and Majorons can have two types of couplings, {\it scalar} ($g_s$) or {\it pseudoscalar} ($g_p$):
\begin{equation}
\mathcal{L}_{\rm int}=\frac{(g_s)_{ij}}{2}\bar\nu_i\nu_jJ+i\frac{(g_p)_{ij}}{2}\bar\nu_i\gamma_5\nu_j J~.
\end{equation}
The decay widths generated by such couplings are well known and can be found, for instance, in~\cite{Lindner:2001fx, Gago:2017zzy}. Relevant constraints to these couplings can be found in the literature~\cite{Kachelriess:2000qc, Pasquini:2015fjv, Gando:2012pj, Agostini:2015nwa, Hannestad:2005ex, Archidiacono:2013dua}.

In general, most works consider that the light neutrinos can decay into an unobservable product, such as a light sterile neutrino or an undetectable active neutrino. This is the so-called {\it invisible decay} (ID). However, the possibility of observing the decay into active products, referred to as {\it visible decay} (VD), has also been studied in the past~\cite{Lindner:2001fx}, and has recently been reconsidered in the context of current long-baseline experiments~\cite{Gago:2017zzy}. The trait of this scenario is an excess of neutrinos of lower energies, which are the decay products of the neutrinos from the original flux. This was followed by an analysis in the framework of future facilities~\cite{Coloma:2017zpg}, which included matter effects. The question of how to properly include matter effects within the oscillation with decay scenario was also addressed in~\cite{Moss:2017pur}, where they considered the decay of a heavier sterile neutrino.

In this paper we take a different approach in the inclusion of matter effects, following closely the prescription outlined in~\cite{Lindner:2001fx}. The role of matter in the visible component of the decay process is analyzed by comparing two different baselines (corresponding to DUNE~\cite{Acciarri:2015uup} and ANDES~\cite{Bertou:2012fk}), which correspond to different matter densities. In addition, we make a systematic study of the sensitivity to the decay parameters at the DUNE experiment, including all oscillation channels and the two modes of operation (FHC and RHC).

This paper goes as follows. On Section~\ref{sec:newformula}, we develop the theoretical framework to be used for describing neutrino decay in matter. On Section~\ref{sec:nevents}, we assess the impact of matter effects in neutrino decay within the aforementioned baselines. Finally, on Section~\ref{sec:paramfits} we discuss the details of our DUNE simulation and statistical analysis and evaluate the sensitivity of this experiment to the decay constant $\alpha_3$. On that Section we also study the impact of neutrino decay on oscillation fits.
 
\section{Neutrino flux including matter effects and decay}
\label{sec:newformula}

The main idea behind our procedure relies on carefully identifying the basis where each process takes place. As is common, one defines the {\it interaction} (or flavour) basis as the one where the charged lepton mass matrix is diagonal, which is also the basis where neutrino interaction states ($\nu_e$, $\nu_\mu$, $\nu_\tau$) are produced. However, on this basis the neutrino mass matrix is not diagonal, meaning that the interaction states are a superposition of mass eigenstates. This mass matrix is diagonalized by the PMNS matrix $(U_0)_{\alpha i}$, where $\alpha=e,\,\mu,\,\tau$, and $i=1,2,3$, which connects both states. We refer to the basis where the neutrino mass matrix is diagonal as the {\it mass} basis, and it is in this basis where the decay widths $\Gamma_i$ of each neutrino are defined.

As is well known, when neutrinos travel through matter, the effective Hamiltonian $H$ is not diagonal, neither in the interaction nor mass bases. For our case, in terms of $\alpha_i=E\, \Gamma_i$, we have on the interaction basis:
\begin{equation}
 H=\frac{1}{2E}\,U_0\left(\begin{array}{ccc}
m_1^2 & & \\ & m_2^2-i\,\alpha_2 & \\ & & m_3^3-i\,\alpha_3 
\end{array}\right)U_0^\dagger+\left(\begin{array}{ccc}
\sqrt{2}G_F N_e & & \\ & 0 & \\ & & 0
\end{array}\right)~,
\end{equation}
where $E$ is the neutrino energy, $G_F$ is Fermi's constant, and $N_e$ is the electron density in matter. We have assumed normal ordering with the lightest neutrino being stable.

This structure motivates the introduction of a new basis~\cite{Wolfenstein:1977ue,Mikheev:1986gs}, which we refer to as the {\it matter} basis, where the effective Hamiltonian is diagonal. Given the presence of the decay widths, $H$ is diagonalized by non-unitary $\tilde U_{\alpha I}$ matrices ($I=\tilde 1,\,\tilde 2,\,\tilde 3$):
\begin{equation}
\label{eq:mixing}
\tilde U^{-1}H\, \tilde U = H^{\rm diag}~,
\end{equation}
where $H^{\rm diag}$ has complex eigenvalues:
\begin{equation}
\label{eq:eigenvalues}
\tilde m_I^2-i\,\tilde\alpha_I=2E\,(H^{\rm diag})_{II}
\end{equation}
It is important to take into account that, in this new basis, all eigenstates can have a non-zero decay width. For instance, if we start with only $\alpha_3$ different from zero, we can obtain non-vanishing $\tilde\alpha_1$, $\tilde\alpha_2$ and $\tilde\alpha_3$.

Let us now understand how the combination of matter effects and decay affect a neutrino flux. We denote the flux arriving at the detector in absence of flavour transitions by $d\Phi_\alpha^{(r)}/dE_{\alpha}$, where $\alpha$ refers to the flavour, $r=(+,\,-)$ is the helicity, and $E_\alpha$ is the energy, which are determined when the neutrino is produced. In the presence of oscillations and decay, the neutrino flux of flavour $\beta$, helicity $s$ and energy $E_\beta$, can be calculated with:
\begin{equation}
 \label{eq:flux}
 \frac{d \Phi^{(s)}_\beta}{dE_\beta}=\int P_{\rm dec}\left(\nu_\alpha^{(r)}\to\nu_\beta^{(s)}\right)\frac{d\Phi^{(r)}_\alpha}{dE_\alpha}dE_\alpha
\end{equation}
The transition function $P_{\rm dec}(\nu_\alpha^{(r)}\to\nu_\beta^{(s)})$ is~\cite{PalomaresRuiz:2005vf,Gago:2017zzy}:
\begin{eqnarray}
\label{eq:transition}
P_{\rm dec}\left(\nu_\alpha^{(r)}\to\nu_\beta^{(s)}\right)&=&
\left|\sum_{I=1}^3\left(\tilde U^{(r)}\right)_{I\alpha}^{-1}\exp\left[-i\frac{\tilde m_I^2 L}{2E_\alpha}\right]\exp\left[-\frac{\tilde \alpha_I L}{2E_\alpha}\right]\tilde U_{\beta I}^{(s)}\right|^2
\delta_{rs}\,\delta(E_\alpha-E_\beta) \nonumber \\
&&+P_{\rm vis}(E_\alpha,\,E_\beta)
\end{eqnarray}
The first term gives the ID contribution, which also includes neutrino oscillations. This term describes the oscillated neutrino flux, taking into account the loss of neutrinos due to the decay. The surviving neutrinos do not change their energy nor helicity, so $\delta(E_\alpha-E_\beta)$ and $\delta_{rs}$ terms need to be added. Notice that the mixing depends on the helicity: $\tilde U^{(-)}=\tilde U$ and $\tilde U^{(+)}=\tilde U^*$.

As mentioned in the Introduction, in this work we assume that the neutrinos from the original flux decay into lighter ones. The decay products, for all practical purposes, have the same direction as the initial neutrinos, and reach the far detector at the same time. Thus, they become a secondary contribution to the neutrino flux. This secondary contribution is described by the second term of Eq.~(\ref{eq:transition}), and is referred to as VD.

One important thing is that VD does not have a $\delta(E_\alpha-E_\beta)$ function, so the state after the decay can have a different energy compared to the one before the decay. As the effective Hamiltonian depends on the energy, this means that the eigenvalues and mixing (Eqs.~(\ref{eq:mixing}) and~(\ref{eq:eigenvalues})) before the decay shall be different to those after the decay. Thus, we denote the matter basis before the decay with a tilde, and the matter basis after the decay with a hat (for example, $\tilde m$ vs $\hat m$).

We refer to the addition of ID and VD contributions as {\it full decay} (FD). In contrast, the results without neutrino decay are referred to as {\it standard oscillations} (SO).

The matrices $\tilde U$ and $\hat U$ relate the interaction eigenstates $\alpha,\,\beta$ with the matter eigenstates $I,\,J$, which is important to connect the neutrino production and detection with its propagation. In the same way, to connect the neutrino propagation with its decay, we need the matrices $\tilde C$ and $\hat C$, that relate the matter eigenstates $I,\,J$ with the mass eigenstates $i,\,j$. We define them in the following way:
\begin{align}
 \tilde C_{Ij}^{(r)}=\sum_{\rho=e,\mu,\tau}\tilde U_{\rho I}^{(r)}(U_0)_{\rho j}^{(r)*} & &
 \hat C_{Ij}^{(s)}=\sum_{\rho=e,\mu,\tau}\hat U_{\rho I}^{(s)}(U_0)_{\rho j}^{(s)*}
\end{align}

With this notation, assuming that the neutrino decays only once in its path, we can use the methods described in~\cite{Lindner:2001fx} to write:
\begin{eqnarray}
\label{eq:pvis_basic}
P_{\rm vis}(E_\alpha,\,E_\beta)&=& \int d\ell\,\left|\sum_{I=\tilde 1}^{\tilde 3}\left(\tilde U^{(r)}\right)^{-1}_{I\alpha} 
\exp\left[-i\frac{\tilde m_I^2\, \ell}{2E_\alpha}\right]\exp\left[-\frac{\tilde \alpha_I\, \ell}{2E_\alpha}\right]
\sum_{i=2}^3\sum_{j=1}^{i-1}\tilde C^{(r)}_{Ii}\sqrt{\frac{d}{dE_\beta}\Gamma_{\nu_i^r\to\nu_j^s}(E_\alpha)}\right. \nonumber \\
&&\left.\times\sum_{J=\hat 1}^{\hat 3}\left(\hat C^{(s)}\right)_{jJ}^{-1}\exp\left[-i\frac{\hat m_J^2 (L-\ell)}{2E_\beta}\right]\exp\left[-\frac{\hat \alpha_J (L-\ell)}{2E_\beta}\right]\hat U_{\beta J}^{(s)}\right|^2
\end{eqnarray}

The above equation can be understood as follows: first, the source generates a neutrino interaction eigenstate, described by the index $\alpha$. The propagation occurs on the basis where $H(E_\alpha)$ is diagonal, so we need to use $\tilde U^{-1}$ to rotate into the matter eigenstates, with index $I$. These states propagate a distance $\ell$, and then decay. The decay, however, is defined on the mass basis in vacuum, so we switch to this basis (index $i$) using $\tilde C$. After the decay, one has a mass eigenstate $\nu_j^{(s)}$ that must be propagated a distance $(L-\ell)$. Nevertheless, this state is again not an eigenstate of $H(E_\beta)$, meaning we need to change basis again. For this, we use $\hat C^{-1}$ to rotate into matter eigenstates (index $J$). These states propagate the distance $(L-\ell)$, and then interact with the detector. One obtains the final flavour $\beta$ by switching back to the interaction basis using $\hat U$.

Assuming constant $N_e$, one can evaluate Eq.~(\ref{eq:pvis_basic}) using~\cite{Lindner:2001fx}. We obtain:
\begin{eqnarray}
 P_{\rm vis}(E_\alpha,\,E_\beta)&=& 2\sum_{I=\tilde 1}^{\tilde 3}\sum_{J=\hat 1}^{\hat 3}\sum_{M=\tilde 1}^{\tilde 3}\sum_{N=\hat 1}^{\hat 3}
 \left(\tilde U^{(r)}\right)_{I \alpha}^{-1}\left(\tilde U^{(r)}\right)_{M\alpha}^{-1*}\,\hat U_{\beta J}^{(s)}\,\hat U_{\beta N}^{(s)*} \nonumber \\
 &&\times E_\beta\frac{\left[(E_\beta/E_\alpha)\tilde\alpha_{<IM>}-\hat\alpha_{<JN>}\right]-i\left[(E_\beta/E_\alpha)\Delta \tilde m^2_{IM}-\Delta \hat m^2_{JN}\right]}{\left[(E_\beta/E_\alpha)\tilde\alpha_{<IM>}-\hat\alpha_{<JN>}\right]^2+\left[(E_\beta/E_\alpha)\Delta \tilde m^2_{IM}-\Delta \hat m^2_{JN}\right]^2} \nonumber \\
 &&\times\left\{\exp\left[-i\frac{\Delta\hat m_{JN}^2 L}{2E_\beta}\right]\exp\left[-\frac{\hat\alpha_{<JN>}L}{2E_\beta} \right]-\exp\left[-i\frac{\Delta\tilde m_{IM}^2 L}{2E_\alpha}\right]\exp\left[-\frac{\tilde\alpha_{<IM>}L}{2E_\alpha} \right]\right\}\nonumber \\
 &&\times\sum_{i=2}^3\sum_{j=1}^{i-1}\sum_{m=2}^3\sum_{n=1}^{m-1}
 \tilde C_{Ii}^{(r)}\tilde C_{Mm}^{(r)*}\left(\hat C^{(s)}\right)_{jJ}^{-1}\left(\hat C^{(s)}\right)_{nN}^{-1*}
 \sqrt{\frac{d}{dE_\beta}\Gamma_{\nu_i^r\to\nu_j^s}(E_\alpha)\frac{d}{dE_\beta}\Gamma_{\nu_m^r\to\nu_n^s}(E_\alpha)} \nonumber \\
\end{eqnarray}
where we have generically denoted $\alpha_{<IJ>}=\alpha_I+\alpha_J$. The calculation of $d\Gamma(\nu_{\rm ini}^r\to\nu_{\rm fin}^s\,J)/dE_\beta$ on the mass basis has been done before~\cite{Gago:2017zzy}.

In this work, we shall again consider only one decay channel $(\nu^r_3\to\nu^s_1 J)$ and only one non-vanishing coupling. In this limit, we find our final expression:
\begin{eqnarray}
 \label{complete-equation}
 P_{\rm vis}(E_\alpha,\,E_\beta)&=& 2\sum_{I=1}^3\sum_{J=1}^{3}\sum_{M=1}^3\sum_{N=1}^3
 \left(\tilde U^{(r)}\right)_{I \alpha}^{-1}\left(\tilde U^{(r)}\right)_{M\alpha}^{-1*}\hat U_{\beta J}^{(s)}\hat U_{\beta N}^{(s)*} \nonumber \\
 &&\times \frac{\left[(E_\beta/E_\alpha)\tilde\alpha_{<IM>}-\hat\alpha_{<JN>}\right]-i\left[(E_\beta/E_\alpha)\Delta \tilde m^2_{IM}-\Delta \hat m^2_{JN}\right]}{\left[(E_\beta/E_\alpha)\tilde\alpha_{<IM>}-\hat\alpha_{<JN>}\right]^2+\left[(E_\beta/E_\alpha)\Delta \tilde m^2_{IM}-\Delta \hat m^2_{JN}\right]^2} \nonumber \\
 &&\times\left\{\exp\left[-i\frac{\Delta\hat m_{JN}^2 L}{2E_\beta}\right]\exp\left[-\frac{\hat\alpha_{<JN>}L}{2E_\beta} \right]-\exp\left[-i\frac{\Delta\tilde m_{IM}^2 L}{2E_\alpha}\right]\exp\left[-\frac{\tilde\alpha_{<IM>}L}{2E_\alpha} \right]\right\}\nonumber \\
 &&\times
 \tilde C_{I3}^{(r)}\tilde C_{M3}^{(r)*}\left(\hat C^{(s)}\right)_{1J}^{-1}\left(\hat C^{(s)}\right)_{1N}^{-1*}
 \left(\frac{(E_\beta/E_\alpha)\,\alpha_3}{E_\alpha}\right)\left(1-\frac{m_3^2}{E_\alpha^2}\right)^{-1/2}\frac{x_{31}^2}{(x_{31}^2-1)}F^{rs}_g(E_\alpha,\,E_\beta) \nonumber \\
 &&\times\Theta_H(E_\alpha-E_\beta)\,\Theta_H(x_{31}^2E_\beta-E_\alpha)
\end{eqnarray}
where $g=\{g_s,\,g_p\}$ indicates the non-vanishing coupling, $x_{if}=m_i/m_f>1$, $\Theta_H(x)$ is the Heaviside function, and:
\begin{subequations}
\label{eq:BigF}
\begin{align}
 F^{\pm\pm}_{g_s}(E_\alpha,\,E_\beta) &= \frac{1}{E_\alpha\,E_\beta}\frac{(E_\alpha+x_{if}E_\beta)^2}{(x_{if}+1)^2}  &
 F^{\pm\mp}_{g_s}(E_\alpha,\,E_\beta) &= \frac{(E_\alpha-E_\beta)}{E_\alpha\,E_\beta}\frac{(x_{if}^2E_\beta-E_\alpha)}{(x_{if}+1)^2}  \\
 F^{\pm\pm}_{g_p}(E_\alpha,\,E_\beta) &= \frac{1}{E_\alpha\,E_\beta}\frac{(E_\alpha-x_{if}E_\beta)^2}{(x_{if}-1)^2} &
 F^{\pm\mp}_{g_p}(E_\alpha,\,E_\beta) &= \frac{(E_\alpha-E_\beta)}{E_\alpha\,E_\beta}\frac{(x_{if}^2E_\beta-E_\alpha)}{(x_{if}-1)^2}
\end{align}
\end{subequations}
The reader should be aware that in the next-to-last line of Eq.~(\ref{complete-equation}), we have the plain $\alpha_3$ on the mass basis in vacuum. This represents the neutrino-Majoron couplings.

We point out that in~\cite{Gago:2017zzy} it was shown that the scalar and pseudoscalar couplings give undistinguishable effects when the $\nu_1$ mass, $m_{\rm lightest}$, is vanishing. In contrast, when $m_{\rm lightest}=0.07$~eV, its largest value allowed by cosmology~\cite{Ade:2013zuv}, the choice of coupling leads to a different phenomenology. From now on, we shall write $x_{31}\to\infty$ when $m_{\rm lightest}$ is vanishing, and $x_{31}\to1$ when $m_{\rm lightest}=0.07$~eV.

An interesting feature of Eq.~(\ref{complete-equation}) is that one could have oscillations not only before, but also after the decay, even in the case of only one decay channel. However, we have checked that, for current neutrino beam energies, this would occur for values of $N_e$ ten times larger than those found on Earth. Thus, we do not pursue this effect any further.

\section{Impact on $(\Phi\times\sigma)$ at DUNE and ANDES}
\label{sec:nevents}

As a first approach in our analysis, we characterize the spectrum of the flux, weighted by its corresponding cross-section. We use:
\begin{equation}
\label{eq:phisigma}
(\Phi\times\sigma)_\beta\equiv
\sum_{s}\sigma^{s,\,\rm CC}_\beta(E_\beta)\frac{d\Phi^{(s)}_\beta}{dE_\beta}~,
\end{equation}
where $d\Phi^{(s)}_\beta/dE_\beta$ was defined in Eq.~(\ref{eq:flux}) and $\sigma_\beta^{s,\,\rm CC}$ is the charged-current cross-section for a neutrino with helicity $s$ and flavour $\beta$. This has been evaluated according to the signal channels in the AEDL rules presented in Section~\ref{sec:paramfits} (Table~\ref{tab:AEDL_Rules}), implying that this parameter encodes both contributions from helicity-conserving and helicity-changing VD.

We are interested in experiments where matter effects can be relevant in neutrino decay. We shall consider DUNE~\cite{Acciarri:2015uup}, and a future hypothetical experiment based on the Agua Negra Deep Experiment Site (ANDES) underground laboratory~\cite{Bertou:2012fk}.

DUNE is one of the most promising experiments in neutrino physics, and is foreseen to begin operation within the next ten years. It comprises a 1300~km baseline, starting at the Long-Baseline Neutrino Facility (LBNF) at Fermilab, and extending to Sanford Underground Research Facility (SURF). This corresponds to an average matter density of $\rho_{\rm DUNE}=2.96$~g/cm$^3$. The beam will be generated using a 1.2~MW primary proton beam from the Main Injector, and is expected to be detected using a massive liquid argon time-projection chamber (LArTPC) detector, located deep underground at SURF. Among other goals, the DUNE experiment aims to measure the CP violating phase $\delta_{\rm CP}$, determine the neutrino mass ordering, resolve the octant for the atmospheric mixing angle, search for proton decay and detect and measure the $\nu_e$ flux from a core-collapse supernova within our galaxy~\cite{Acciarri:2015uup}.

The neutrino flux at the Far Detector used in our simulation was provided by the DUNE collaboration~\cite{Alion:2016uaj}, in both neutrino mode (Forward Horn Current - FHC) and antineutrino mode (Reverse Horn Current - RHC). We assume $1.47\times10^{21}$ protons-on-target (POT) per year~\cite{Acciarri:2015uup,Alion:2016uaj,Strait:2016mof}, for $3.5$ years on each mode. The 40 kt Far Detector will consist of four LArTPC modules~\cite{Acciarri:2016ooe}, which provide a fine-grained image of neutrino scattering events. The latter information shall be relevant in our full simulation, to be presented on Section~\ref{sec:paramfits}.

On the other hand, ANDES is a proposed underground laboratory in the Southern Hemisphere. It would be built in the deepest part ($\sim$1750 m) of the Agua Negra tunnel, linking Argentina and Chile below the Andes mountain range~\cite{Bertou:2012fk, Machado:2012ee}. Its construction, together with the tunnel, is planned to happen around 2018-2026. The topics of the ANDES scientific program include neutrinos, dark matter, geophysics, biology, among others~\cite{Last}.

With the aim of evaluating Eq.~(\ref{eq:phisigma}) in the context of ANDES, we assume a hypothetical neutrino beam starting at Fermilab. The corresponding neutrino flux is identical to the one we use for DUNE, but properly scaled to match the distance from source to detector. The ANDES baseline would be of order 7650~km~\cite{andespage}, which corresponds\footnote{This value was calculated using the model of the density profile of the Earth~\cite{Dziewonski:1981xy}, the distance between Fermilab and ANDES, and the radius of the Earth.} to an average matter density $\rho_{\rm ANDES}=4.7$~g/cm$^3$. Thus, we can expect matter effects in neutrino decay to be much more important than in DUNE.

\begin{figure}[tb]
\centering
\includegraphics[width=0.45\textwidth]{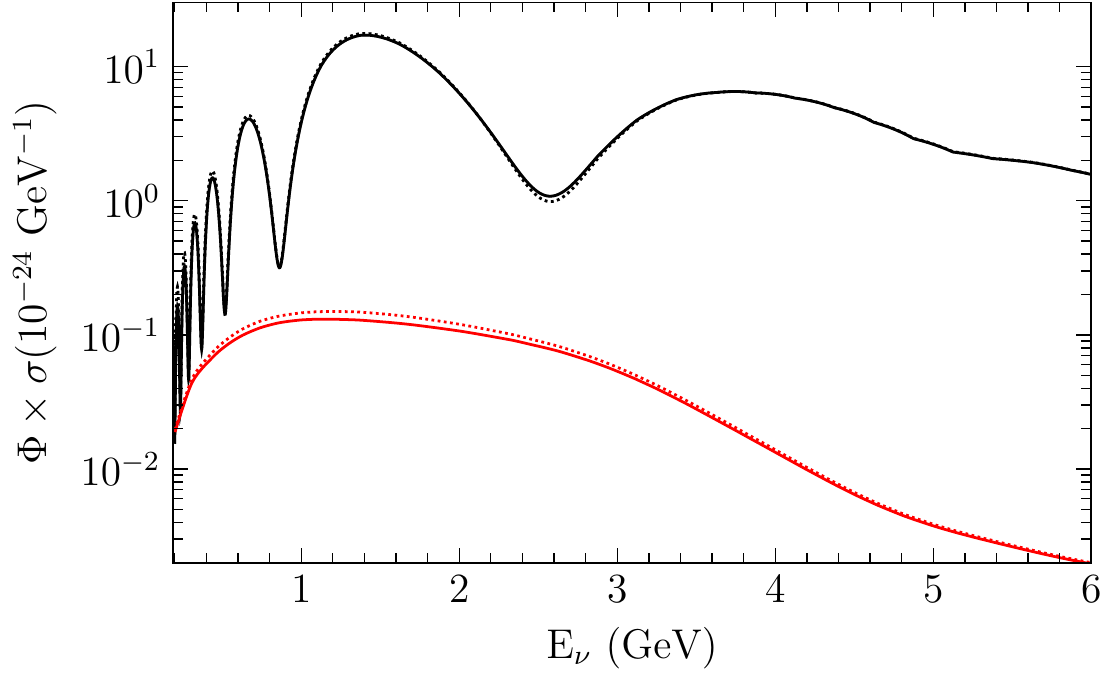} \hfill
\includegraphics[width=0.45\textwidth]{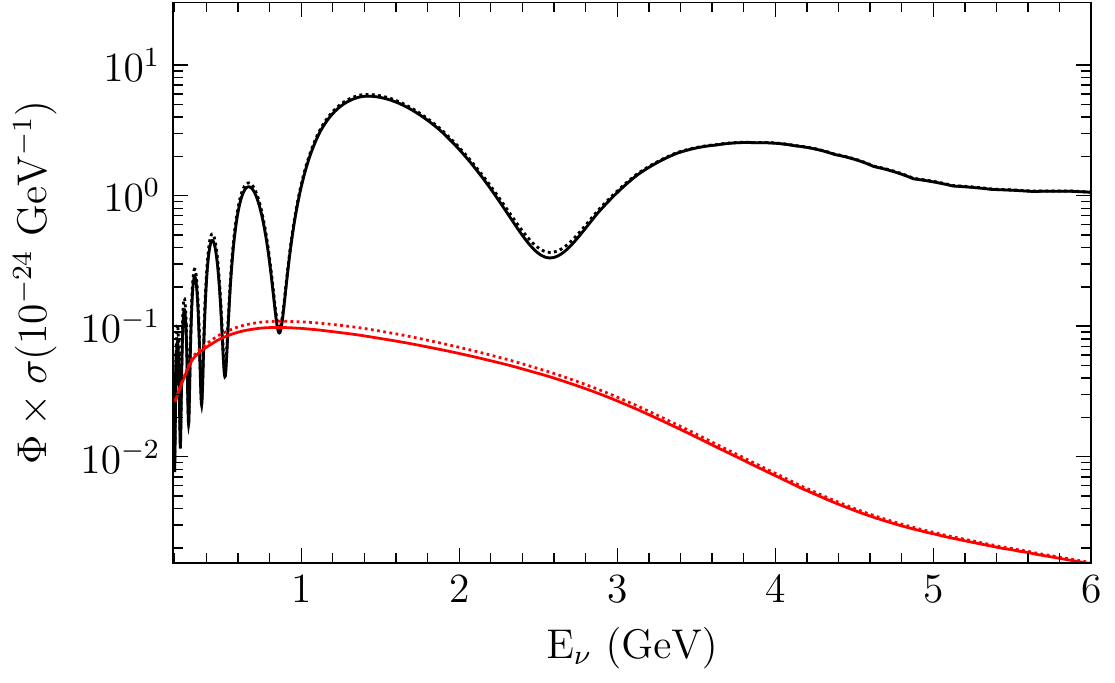} \\
\includegraphics[width=0.45\textwidth]{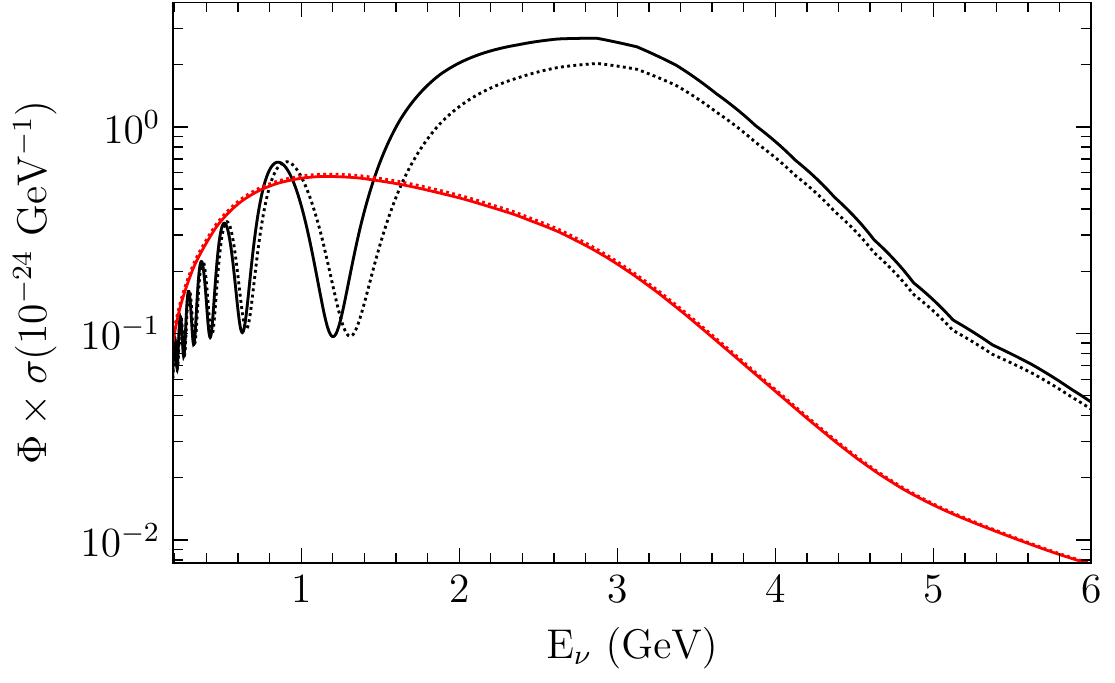} \hfill
\includegraphics[width=0.45\textwidth]{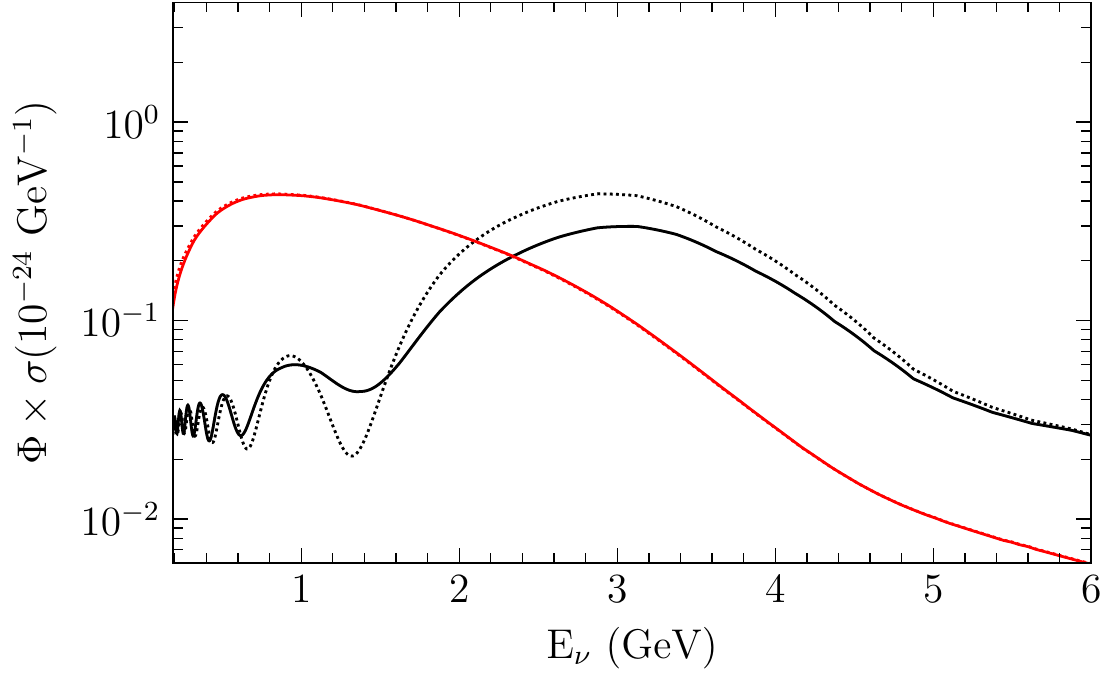} 
\caption{$(\Phi\times\sigma)$ at the DUNE baseline, for $\nu_\mu$ disappearance (top) and $\nu_e$ appearance (bottom) channels. The left (right) column shows results considering the FHC (RHC) flux. The black and red curves corresponds to ID and VD, respectively. Solid lines consider matter effects, while dashed lines show an equivalent scenario in vacuum.}
\label{fig:phisig.DUNE}
\end{figure}

We fix the neutrino mass differences and mixing parameters at the following values~\cite{Esteban:2016qun}: $s^2_{12}=0.306$, $s^2_{23}=0.441$, $s^2_{13}=0.02166$, $\Delta m^2_{21}=7.5\times10^{-5}$~eV$^2$ and $\Delta m^2_{31}=2.524\times10^{-3}$~eV$^2$, for normal ordering. We also set $\delta_{\rm CP}=-\pi/2$ and $\alpha_3=4\times10^{-5}$~eV$^2$, the latter corresponding roughly to a $10\%$ of $\langle E_\alpha\rangle/L$ for DUNE. For VD, we set $x_{31}\to\infty$, such that the type of coupling becomes irrelevant. For definiteness, from now on we choose only a non-vanishing scalar $(g_s)_{31}$ coupling.

In Figure~\ref{fig:phisig.DUNE} we show the expected $(\Phi\times\sigma)$ for $\nu_\mu$ disappearance and $\nu_e$ appearance at DUNE. We separate the ID and VD components of the flux, in order to properly understand the difference between each contribution. For comparison, we show equivalent curves for the case where no matter effects are present. It is important to note that, even though not shown, for the value of $\alpha_3$ we are using we have confirmed that the ID curves are almost identical to those obtained with the SO hypothesis.

For $\nu_\mu$ disappearance (top row), we find that matter effects are almost inexistent for neither ID nor VD components. The ID contribution dominates both the FHC and RHC modes. Moreover, for FHC we have VD between one and two orders of magnitude lower than ID, while for RHC the difference between contributions is somewhat smaller. We can understand why this ID-VD difference in RHC is smaller than the one for FHC in terms of the helicity-changing decay channel. For FHC, the latter decay channel implies that part of the $\nu^{(-)}$ flux becomes $\nu^{(+)}$. In contrast, on RHC we have a fraction of $\nu^{(+)}$ turning into $\nu^{(-)}$. Since the $\nu^{(-)}$ have a larger cross-section than the $\nu^{(+)}$, the relative ID-VD difference in $(\Phi\times\sigma)$ becomes smaller for RHC. This observation is consistent with our results in~\cite{Gago:2017zzy} for T2K, and shall also be relevant in the $\nu_e$ appearance channel.

This result is strictly valid in the $x_{31}\to\infty$ limit. For larger masses ($x_{31}\to1$), the behaviour depends on the coupling. For a scalar coupling, the helicity-flipping decay channel is suppressed, while for a pseudoscalar coupling it is slightly enhanced. This shall have a direct impact on the VD spectrum.

Regardless of these points, the fact remains that ID dominates the flux. Therefore, since for this value of $\alpha_3$ the ID component coincides with SO, we expect that $\nu_\mu$ disappearance shall not be strongly modified by neutrino decay. Then this channel will constrain $\Delta m^2_{32}$ and $\sin^22\theta_{23}$ reliably, independently of the presence of decay.

For $\nu_e$ appearance, the low value of $\alpha_3$ again implies that the ID spectrum shall be very similar to the one of SO. However, we find that the VD contribution at low energy can dominate the flux, being up to one order of magnitude larger than ID for the RHC flux. Consequently, we can expect a better constraint on $\alpha_3$ by using $\nu_e$ appearance instead of $\nu_\mu$ disappearance data, which is again consistent with the results in~\cite{Gago:2017zzy}. We find that the VD contribution for $\nu_e$ appearance is slightly smaller for FHC than for RHC. As in $\nu_\mu$ disappearance, this can be understood in terms of helicity-flipping decay channels, and the different cross-sections.

\begin{figure}[tb]
\centering
\includegraphics[width=0.45\textwidth]{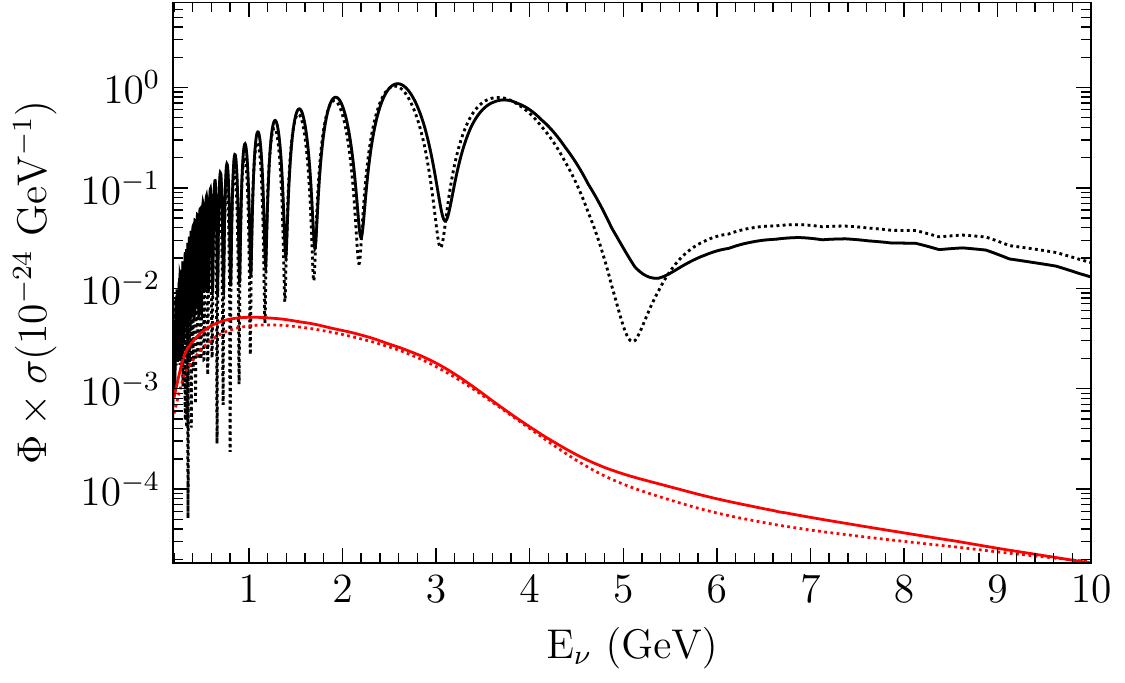} \hfill
\includegraphics[width=0.45\textwidth]{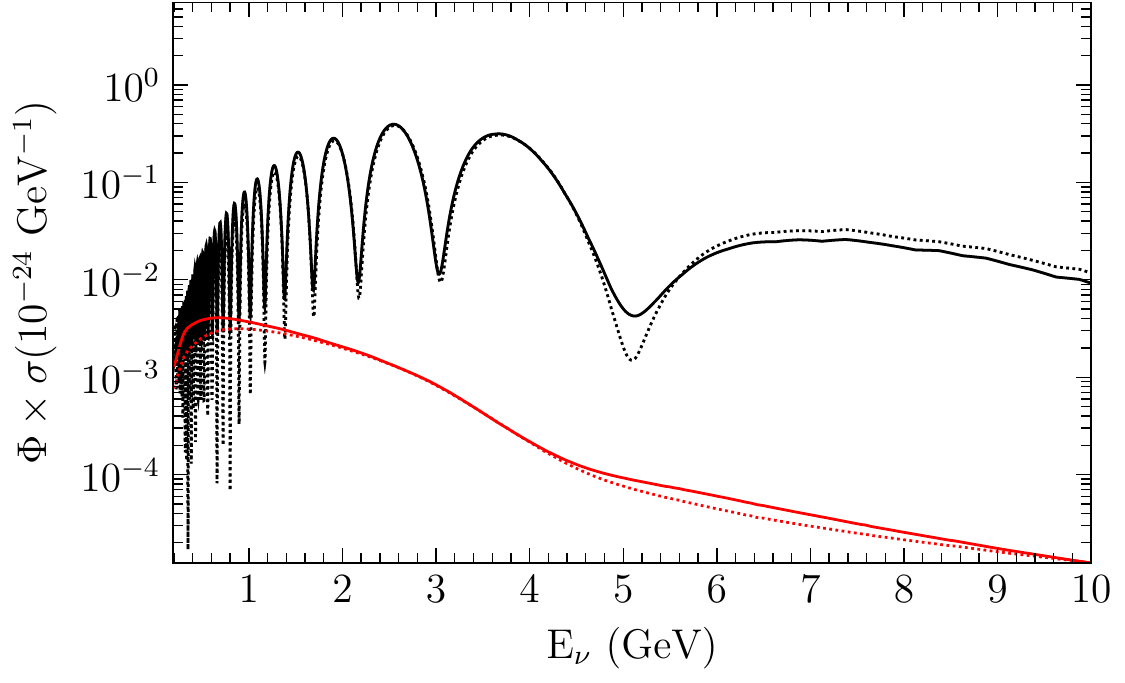} \\
\includegraphics[width=0.45\textwidth]{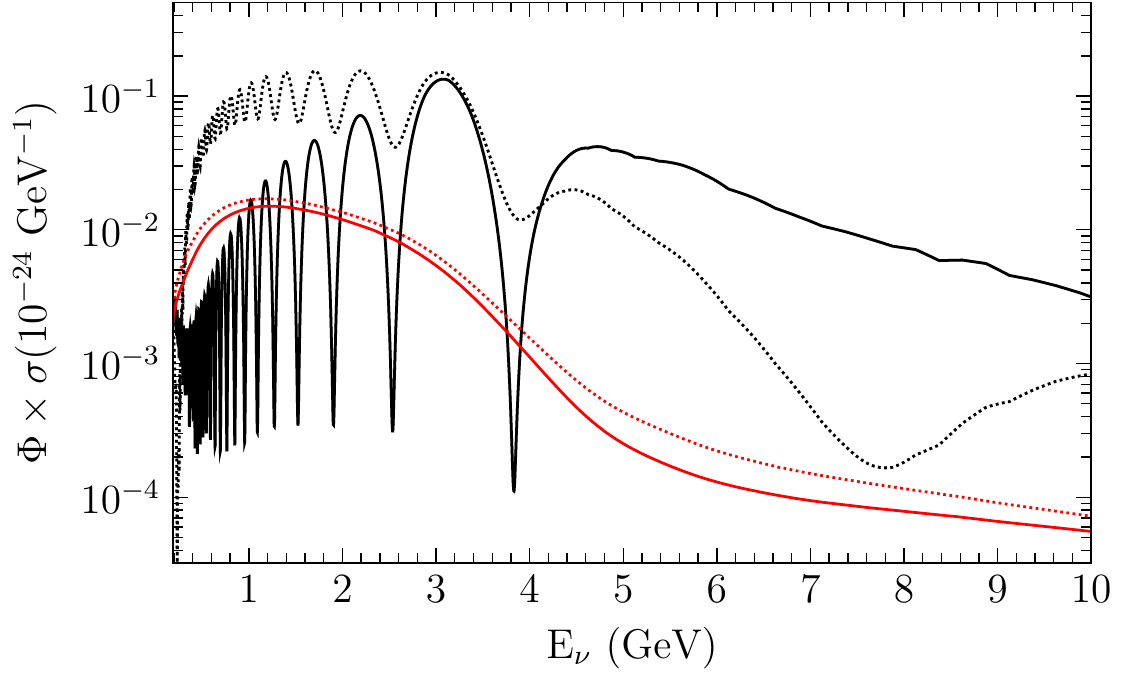} \hfill
\includegraphics[width=0.45\textwidth]{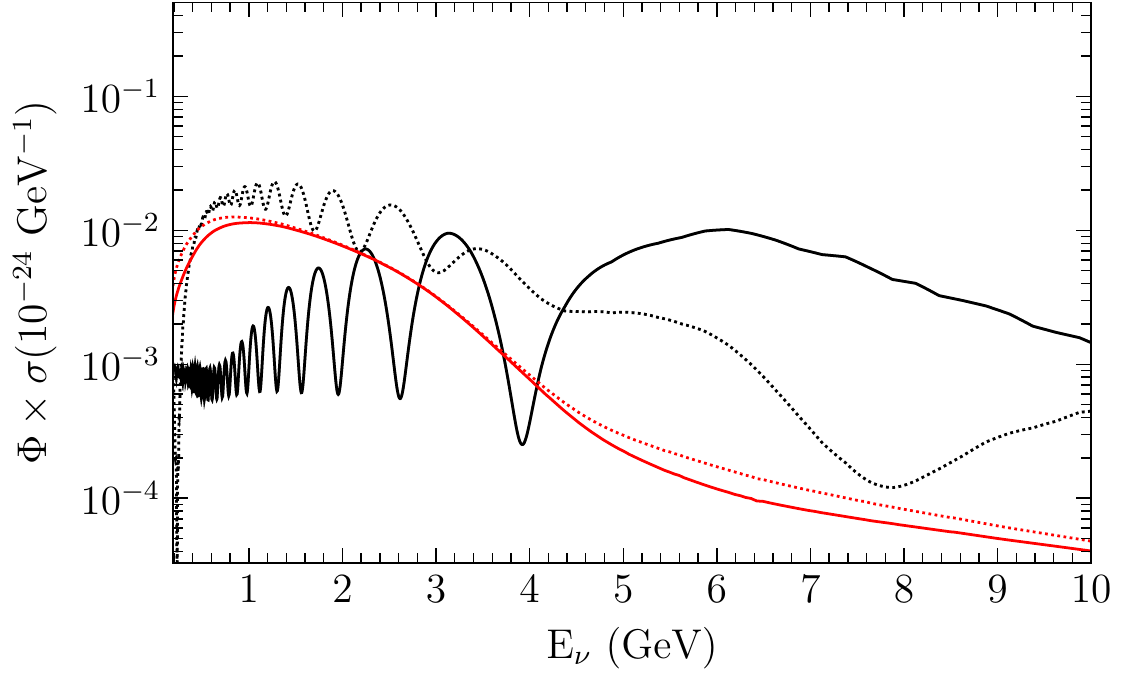} 
\caption{$(\Phi\times\sigma)$ at ANDES, for $\nu_\mu$ disappearance (top) and $\nu_e$ appearance (bottom) channels. The left (right) column shows results considering the FHC (RHC) flux. Curves as in Figure~\ref{fig:phisig.DUNE}.}
\label{fig:phisig.ANDES}
\end{figure}
Another important feature is that matter effects are only relevant for the ID contribution to $\nu_e$ appearance. Given the similarity between ID and SO, these correspond to the typical matter effects for the SO scenario. Thus, a useful result is that, to a very good approximation, one can ignore matter effects in VD completely at this baseline.

In Figure~\ref{fig:phisig.ANDES}, we show both channels for the ANDES baseline. Here, since the baseline is much larger than in DUNE, matter effects are much more relevant. This time, we use $\alpha_3=8\times10^{-6}$~eV$^2$, which again corresponds roughly to $10\%$ of $\langle E\rangle/L$ for ANDES.

A first observation is that, qualitatively, the patterns we observed at DUNE are repeated in ANDES. Nevertheless, quantitatively, matter effects make a difference. It is remarkable that even for ID in $\nu_\mu$ disappearance the discrepancy between vacuum and matter effects is noticeable.

The most striking consequence of the inclusion of matter effects is the suppression of the ID contribution to $\nu_e$ appearance at low energy, leaving a dominant VD component, with respect to the vacuum case. We want to emphasize that even though both DUNE and ANDES feature a large VD contribution in $\nu_e$ appearance, the reason for this in each case is different. In addition, VD decay is affected in more strongly than in DUNE, at most being reduced by $\sim14\%$.

In this channel, the opposite effect happens at larger energy. In vacuum, the difference between ID and VD is smaller than the one in matter. This is due to an enhancement of the ID component in front of matter effects.

\begin{figure}[tb]
\centering
\includegraphics[width=0.45\textwidth]{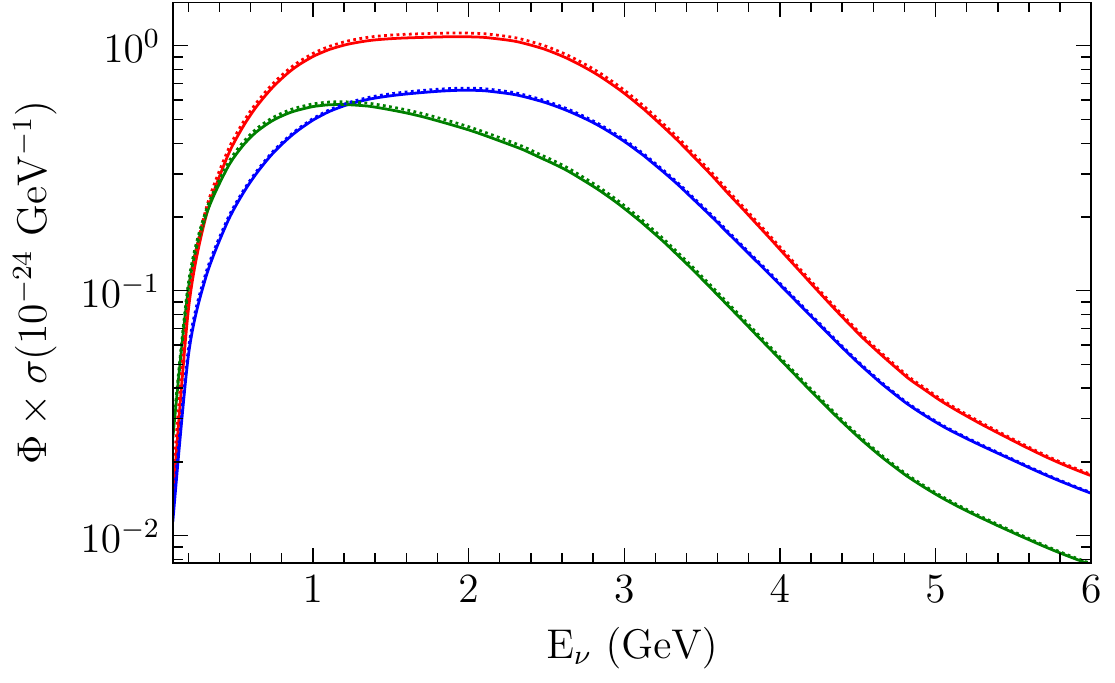} \hfill
\includegraphics[width=0.45\textwidth]{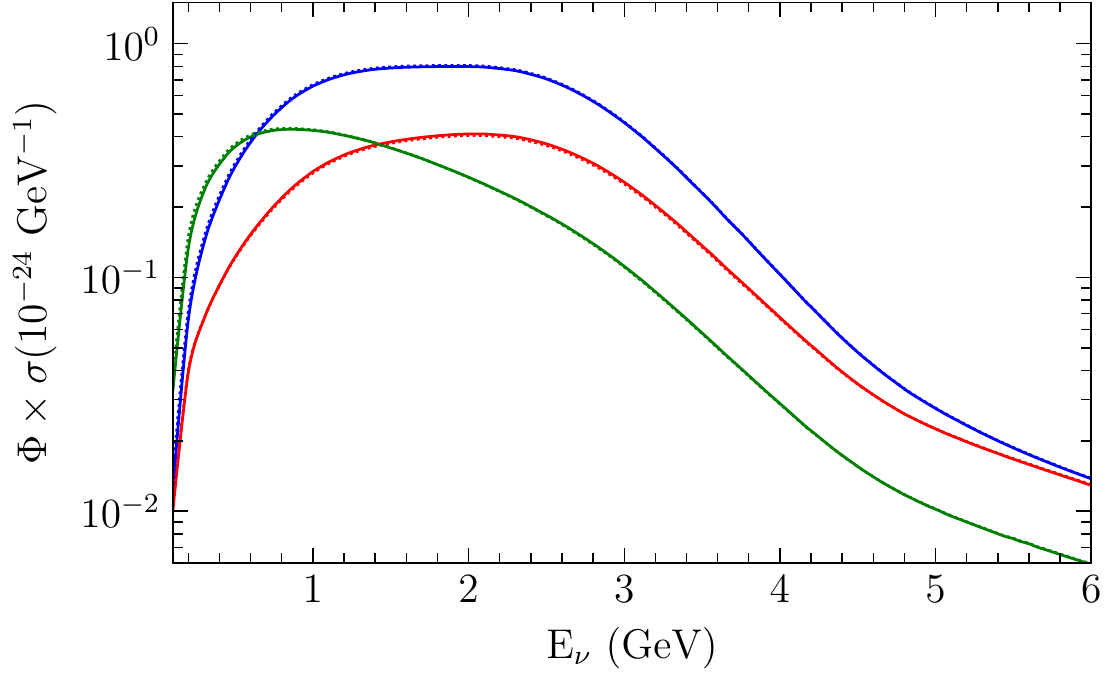} \\
\includegraphics[width=0.45\textwidth]{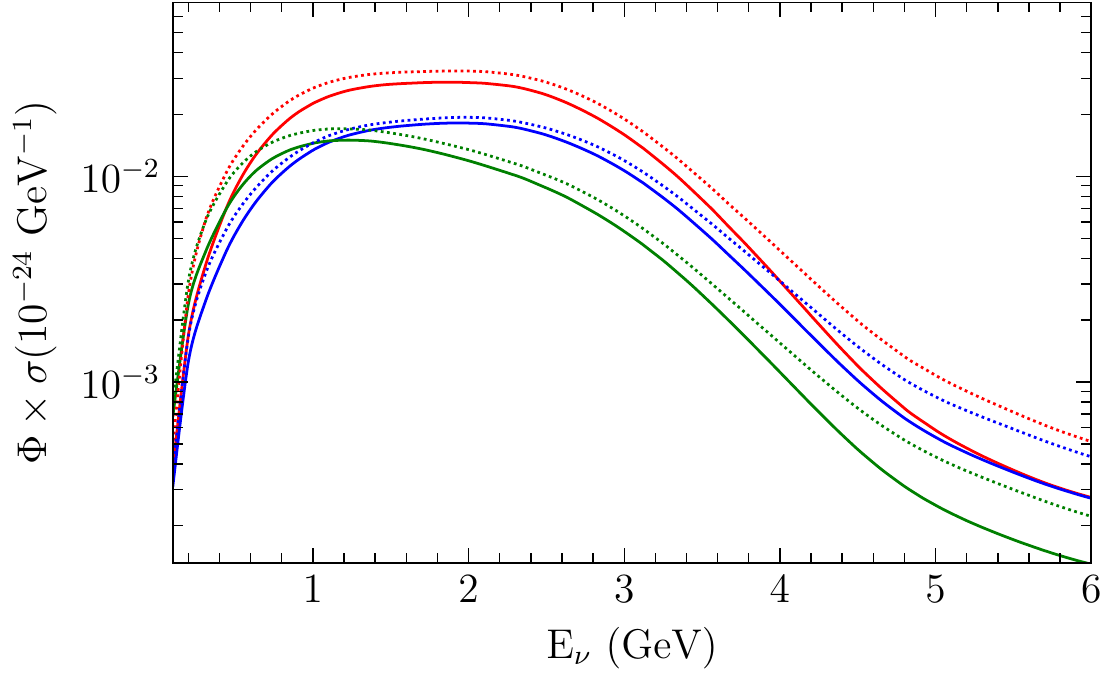} \hfill
\includegraphics[width=0.45\textwidth]{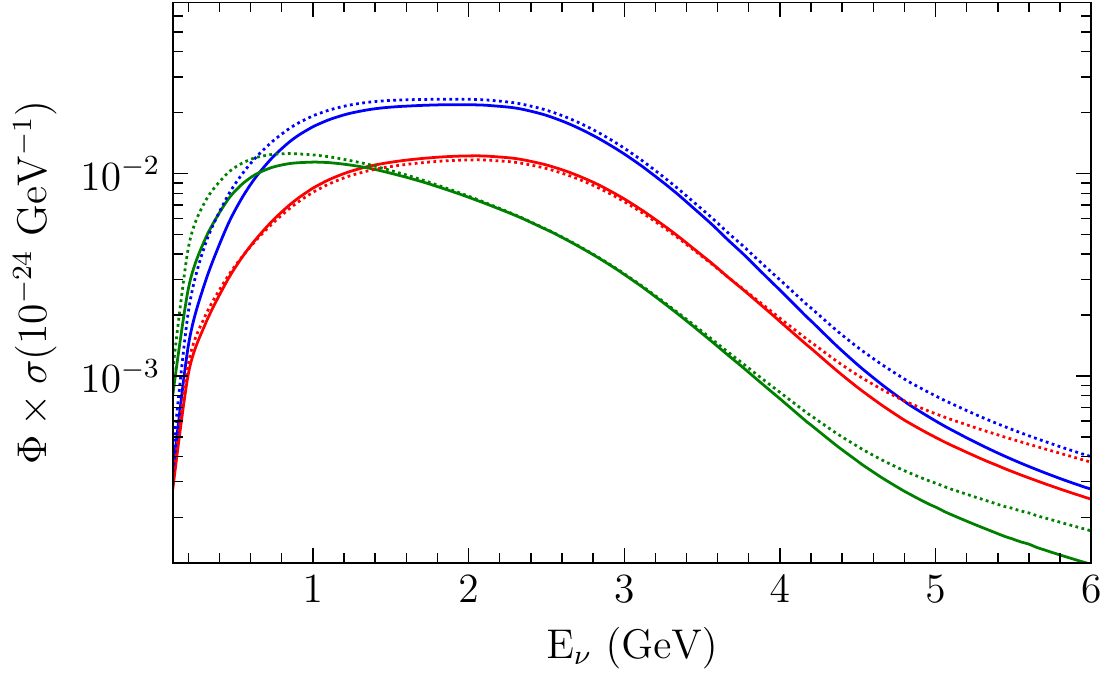} 
\caption{The VD contribution to $(\Phi\times\sigma)$, for $\nu_e$ appearance, at DUNE (top) and ANDES (bottom). The left (right) column shows results considering the FHC (RHC) flux. Solid (dashed) lines include (do not include) matter effects. We show results for $x_{31}\to1$ with scalar and pseudoscalar couplings in red and blue, respectively. The green curve considers $x_{31}\to\infty$, where the two types of coupling give the same effects.}
\label{fig:visible.compare}
\end{figure}
For larger neutrino masses ($x_{31}\to1$), the size and shape of the VD contribution to $(\Phi\times\sigma)$ varies with the type of coupling, as shown in Figure~\ref{fig:visible.compare}. For all plots we have the same qualitative behaviour. For FHC, the curve with $x_{31}\to1$ and scalar coupling is always above the other curves. In contrast, for RHC, it is the curve with $x_{31}\to1$ and pseudoscalar coupling the largest one, with the exception of very small energy, where the curve with $x_{31}\to\infty$ can have higher values. This can be explained in terms of helicity-changing decays, as was discussed earlier.

An important difference between the $x_{31}\to1$ and $x_{31}\to\infty$ scenarios is that the spectrum of the latter favours lower energies. This makes this situation more susceptible to the low energy thresholds of an experiment.

\section{Sensitivity and Parameter Fits at DUNE}
\label{sec:paramfits}

The impact of neutrino decay to the sensitivity to $\alpha_3$ and to parameter fits, at DUNE, has been studied earlier in~\cite{Coloma:2017zpg,Choubey:2017dyu}. Nevertheless, there is still room for further development in these analyses. To begin with, the analysis in~\cite{Choubey:2017dyu} only considered ID, so our results with FD are complementary. With respect to~\cite{Coloma:2017zpg}, which also takes FD into account, we also include the $\nu_\mu$ disappearance channel into our analysis, and display the sensitivity as a function of oscillation parameters. In addition, our approach strictly follows the procedure outlined in~\cite{Lindner:2001fx}, while the formulae used in~\cite{Coloma:2017zpg} seem to use additional assumptions which we do not have (in particular, the arguments used to build Eqs.~(9) and~(13) of that work).

\subsection{Event Generation}

In order to calculate the number of events within an energy bin, we follow the same procedure as in~\cite{Gago:2017zzy}. The number of events of flavour $\beta$ in the energy bin $i$, with helicity $s$ and going through interaction $int=\{CC,\,NC\}$, is obtained from:
\begin{equation}
\label{eq:events}
N^{(s),\rm int}_{i,\beta}=
 \int dE_\beta\,K^{\rm int}_i(E_\beta)\,\sigma^{s,\rm int}_\beta(E_\beta)\frac{d\Phi^{(s)}_\beta}{dE_\beta}~,
\end{equation}
where $\sigma_\beta^{s,\rm int}(E_\beta)$ is the cross section for process $int$, and:
\begin{equation}
\label{eq:kernel}
 K^{\rm int}_i(E_\beta)=\int_{E_{\rm i, min}}^{E_{\rm i, max}}dE_{\rm bin}\,\epsilon^{\rm int}_\beta(E_{\rm bin})\,
 R^{\rm int}(E_{\rm bin}-E_\beta)~.
\end{equation}
The detector efficiency $\epsilon^{\rm int}_\beta(E_{\rm bin})$ and resolution function $R^{\rm int}(E_{\rm bin}-E_\beta)$ are taken from the DUNE Collaboration public files~\cite{Alion:2016uaj}. Both Eq.~(\ref{eq:events}) and~(\ref{eq:kernel}) are used to calculate signal and background events, with $\epsilon^{\rm int}_\beta(E_{\rm bin})$ and $R^{\rm int}(E_{\rm bin}-E_\beta)$ properly adjusted, according to the information in~\cite{Alion:2016uaj}.

For ID, this calculation is carried out within GLoBES~\cite{Huber:2004ka,Huber:2007ji}. However, as the neutrino decay effective Hamiltonian is not Hermitian, we could not use the built-in diagonalization algorithms. Thus, for this case, we modified the source probability library, importing our own probability matrix into GLoBES.

\begin{table}[tb]
\centering
\begin{tabular}{|l|c|c||c|}
\hline
\multicolumn{2}{|c|}{} & $\nu_e$ \textbf{appearance}, FHC Flux & $\bar{\nu}_e$ \textbf{appearance}, RHC Flux \\
\hline 
\multirow{2}{2cm}{Signal} 
& CC: & $(\nu_\mu \rightarrow \nu_e)_{ID}+(\nu_\mu \rightarrow \nu_e)_{VD}$ &
 $(\nu_\mu \rightarrow \nu_e)_{ID}+(\bar\nu_\mu \rightarrow \nu_e)_{VD}$ \\
&  & $+(\bar\nu_\mu \rightarrow \nu_e)_{VD}$ &
 $+(\nu_\mu \rightarrow \nu_e)_{VD}$ \\ 
\cline{2-4}
& CC: & $(\bar{\nu}_\mu \rightarrow \bar{\nu}_e)_{ID}+(\nu_\mu \rightarrow \bar{\nu}_e)_{VD}$ & 
 $(\bar{\nu}_\mu \rightarrow \bar{\nu}_e)_{ID}+(\bar\nu_\mu \rightarrow \bar{\nu}_e)_{VD}$\\
&  & $+(\bar\nu_\mu \rightarrow \bar{\nu}_e)_{VD}$ & 
 $+(\nu_\mu \rightarrow \bar{\nu}_e)_{VD}$\\ \hline
\multirow{8}{2cm}{Background} & CC: & $(\nu_e \rightarrow \nu_e)_{ID}$ & $(\nu_e \rightarrow \nu_e)_{ID}$  \\ \cline{2-4}
& CC: & $(\bar{\nu}_e \rightarrow \bar{\nu}_e)_{ID}$ & $(\bar{\nu}_e \rightarrow \bar{\nu}_e)_{ID}$ \\ \cline{2-4}
& CC: & $(\nu_\mu \rightarrow \nu_\mu)_{ID}+(\nu_\mu \rightarrow \nu_\mu)_{VD}$ & $(\nu_\mu \rightarrow \nu_\mu)_{ID}+(\bar\nu_\mu \rightarrow \nu_\mu)_{VD}$\\ \cline{2-4}
& CC: & $(\bar{\nu}_\mu \rightarrow \bar{\nu}_\mu)_{ID}+(\nu_\mu \rightarrow \bar{\nu}_\mu)_{VD}$ & $(\bar{\nu}_\mu \rightarrow \bar{\nu}_\mu)_{ID}+(\bar\nu_\mu \rightarrow \bar{\nu}_\mu)_{VD}$ \\ \cline{2-4}
& CC: & $(\nu_\mu \rightarrow \nu_\tau)_{ID}+(\nu_\mu \rightarrow \nu_\tau)_{VD}$ & $(\nu_\mu \rightarrow \nu_\tau)_{ID}+(\bar\nu_\mu \rightarrow \nu_\tau)_{VD}$ \\ \cline{2-4}
& CC: & $(\bar{\nu}_\mu \rightarrow \bar{\nu}_\tau)_{ID}+(\nu_\mu \rightarrow \bar{\nu}_\tau)_{VD}$ & $(\bar{\nu}_\mu \rightarrow \bar{\nu}_\tau)_{ID}+(\bar\nu_\mu \rightarrow \bar{\nu}_\tau)_{VD}$\\ \cline{2-4}
& NC: & $(\nu_\mu \rightarrow \nu_\alpha)_{ID}+(\nu_\mu \rightarrow \nu_\alpha)_{VD}$ & $(\nu_\mu \rightarrow \nu_\alpha)_{ID}+(\bar\nu_\mu \rightarrow \nu_\alpha)_{VD}$ \\ \cline{2-4}
& NC: & $(\bar{\nu}_\mu \rightarrow \bar{\nu}_\alpha)_{ID}+(\nu_\mu \rightarrow \bar\nu_\alpha)_{VD}$ & $(\bar{\nu}_\mu \rightarrow \bar{\nu}_\alpha)_{ID}+(\bar\nu_\mu \rightarrow \bar\nu_\alpha)_{VD}$ \\  \hline
\hline
\multicolumn{2}{|c|}{} & $\nu_\mu$ \textbf{disappearance}, FHC Flux & $\bar{\nu}_\mu$ \textbf{disappearance}, RHC Flux\\
\hline 
\multirow{2}{2cm}{Signal} & CC: & $(\nu_\mu \rightarrow \nu_\mu)_{ID}+(\nu_\mu \rightarrow \nu_\mu)_{VD}$ & $(\nu_\mu \rightarrow \nu_\mu)_{ID}+(\bar\nu_\mu \rightarrow \nu_\mu)_{VD}$ \\ \cline{2-4}
& CC: & $(\bar{\nu}_\mu \rightarrow \bar{\nu}_\mu)_{ID}+(\nu_\mu \rightarrow \bar\nu_\mu)_{VD}$ & $(\bar{\nu}_\mu \rightarrow \bar{\nu}_\mu)_{ID}+(\bar\nu_\mu \rightarrow \bar\nu_\mu)_{VD}$\\ \hline
\multirow{4}{2cm}{Background} 
& CC: & $(\nu_\mu \rightarrow \nu_\tau)_{ID}+(\nu_\mu \rightarrow \nu_\tau)_{VD}$ & $(\nu_\mu \rightarrow \nu_\tau)_{ID}+(\bar\nu_\mu \rightarrow \nu_\tau)_{VD}$ \\ \cline{2-4}
& CC: & $(\bar{\nu}_\mu \rightarrow \bar{\nu}_\tau)_{ID}+(\nu_\mu \rightarrow \bar\nu_\tau)_{VD}$ & $(\bar{\nu}_\mu \rightarrow \bar{\nu}_\tau)_{ID}+(\bar\nu_\mu \rightarrow \bar\nu_\tau)_{VD}$ \\ \cline{2-4}
& NC: & $(\nu_\mu \rightarrow \nu_\alpha)_{ID}+(\nu_\mu \rightarrow \nu_\alpha)_{VD}$ & $(\nu_\mu \rightarrow \nu_\alpha)_{ID}+(\bar\nu_\mu \rightarrow \nu_\alpha)_{VD}$ \\ \cline{2-4}
& NC: & $(\bar{\nu}_\mu \rightarrow \bar{\nu}_\alpha)_{ID}+(\nu_\mu \rightarrow \bar\nu_\alpha)_{VD}$ & $(\bar{\nu}_\mu \rightarrow \bar{\nu}_\alpha)_{ID}+(\bar\nu_\mu \rightarrow \bar\nu_\alpha)_{VD}$ \\
\hline
\end{tabular}
\caption{AEDL rules for $\nu_e^{(-)}$, $\nu_e^{(+)}$ appearance, and $\nu_\mu^{(-)}$, $\nu_\mu^{(+)}$ disappearance. We denote $\nu_\alpha=\nu_\alpha^{(-)}$ and $\bar\nu_\alpha=\nu_\alpha^{(+)}$.}
\label{tab:AEDL_Rules}
\end{table}

In the case of VD, we generated the fluxes in Eq.~(\ref{eq:flux}) externally, and used these in Eq.~(\ref{eq:events}), in GLoBES, to calculate the event rates. Finally, we modified the channels of each GLoBES rule, such that helicity change was taken into account. For example, given the channel $\nu_\alpha^{(-)} \rightarrow \nu_\beta^{(-)}$, we add $\nu_\alpha^{(-)} \rightarrow\nu_\beta^{(+)}$.

The full set of rules for each channel, for signal and background, including both ID and VD, is shown in Table~\ref{tab:AEDL_Rules}. The Table separates contributions to signal and background based on the final state. However, they are added in the $\chi^2$ analyses presented below. Notice that for ID, we include the contributions coming from both of the original $\nu_\mu^{(-)}$ and $\nu_\mu^{(+)}$ components of the flux, for both $\nu_\mu$ disappearance and $\nu_e$ appearance channels. This is also done for VD in the $\nu_e$ appearance channel. However, we find it suffices to include only the FHC $\nu_\mu^{(-)}$ (RHC $\nu_\mu^{(+)}$) components in VD for $\nu_\mu$ disappearance, with the other component being negligible.

\subsection{Analysis and Results}

We begin by studying the DUNE sensitivity to $\alpha_3$ in the FD scenario. As $\theta_{13}$ is fixed by reactor $\nu_e$ disappearance measurements~\cite{An:2012eh}, which are not strongly affected by FD~\cite{Gago:2017zzy}, we focus on the effect of varying $\theta_{23}$ and $\delta_{\rm CP}$. As in the previous Section, we present results for $x_{31}\to\infty$, fixing the other oscillation parameters at their best fit values. Data from both $\nu_\mu$ disappearance and $\nu_e$ appearance are taken into account. 

To compute $\chi^2$, we calculate the number of events $N_i$ for each combination of $\theta_{23}$, $\delta_{\rm CP}$ and $\alpha_3$, for the energy bin $i$. $N_i$ is calculated using Eq.~(\ref{eq:events}), adding over helicities and interactions, for signal and backgrounds. This is compared to the events generated by a fixed set of parameters, refered to as \textit{true values}. We define $\chi^2$ using:
\begin{equation}
\chi^2(\theta_{23},\,\delta_{\rm CP},\,\alpha_3,\,\theta^{\rm true}_{23},\,\delta^{\rm true}_{\rm CP},\,\alpha^{\rm true}_3) = \sum_i^{\rm bins} \frac{\left( N_i\left(\theta_{23},\delta_{\rm CP},\alpha_3 \right)- N_i\left(\theta^{\rm true}_{23},\delta^{\rm true}_{\rm CP},\alpha^{\rm true}_3\right) \right)^2}{N_i\left( \theta^{\rm true}_{23},\delta^{\rm true}_{\rm CP},\alpha^{\rm true}_3\right)}
\label{chi2}
\end{equation}
To analyze the sensitivity to $\alpha_3$, we set $\theta_{23}=\theta_{23}^{\rm true}$ and $\delta_{\rm CP}=\delta_{\rm CP}^{\rm true}$. The sensitivity to $\alpha_3$ as a function of $\theta_{23}^{\rm true}$ is obtained by marginalizing over $\delta_{\rm CP}^{\rm true}$, and setting $\alpha_3^{\rm true}=0$~eV$^2$, that is:
\begin{equation}
 \left.\chi^2(\theta^{\rm true}_{23},\,\delta^{\rm true}_{\rm CP},\,\alpha_3,\,\theta^{\rm true}_{23},\,\delta^{\rm true}_{\rm CP},\,0)\right|_{\min \delta^{\rm true}_{\rm CP}}
\end{equation}
Similarly, the sensitivity to $\alpha_3$ as a function of $\delta_{\rm CP}^{\rm true}$ is obtained with:
\begin{equation}
 \left.\chi^2(\theta^{\rm true}_{23},\,\delta^{\rm true}_{\rm CP},\,\alpha_3,\,\theta^{\rm true}_{23},\,\delta^{\rm true}_{\rm CP},\,0)\right|_{\min \theta^{\rm true}_{23}}
\end{equation}

\begin{figure}[tb]
\centering
\includegraphics[width=0.49\textwidth]{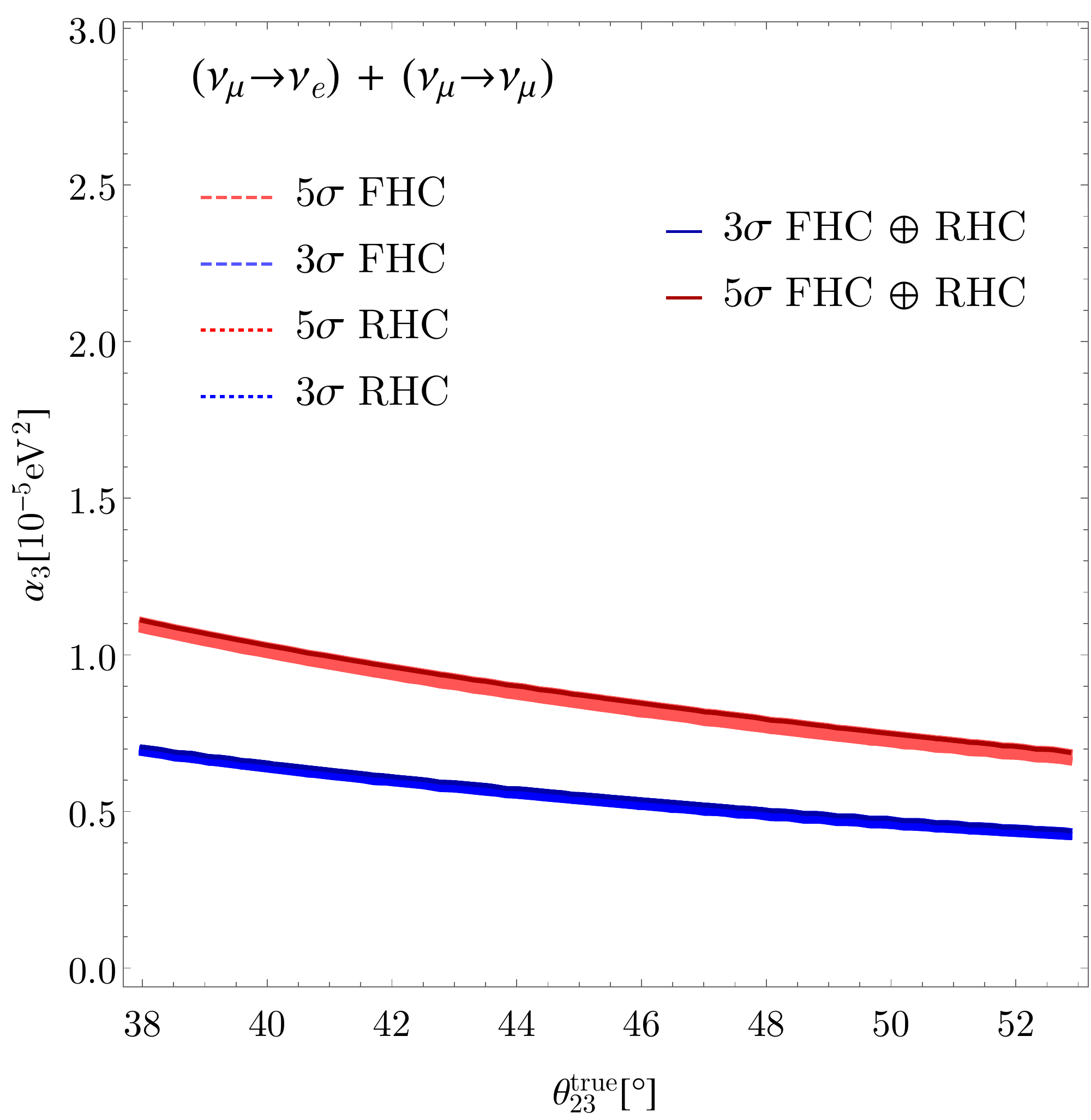} \hfill
\includegraphics[width=0.49\textwidth]{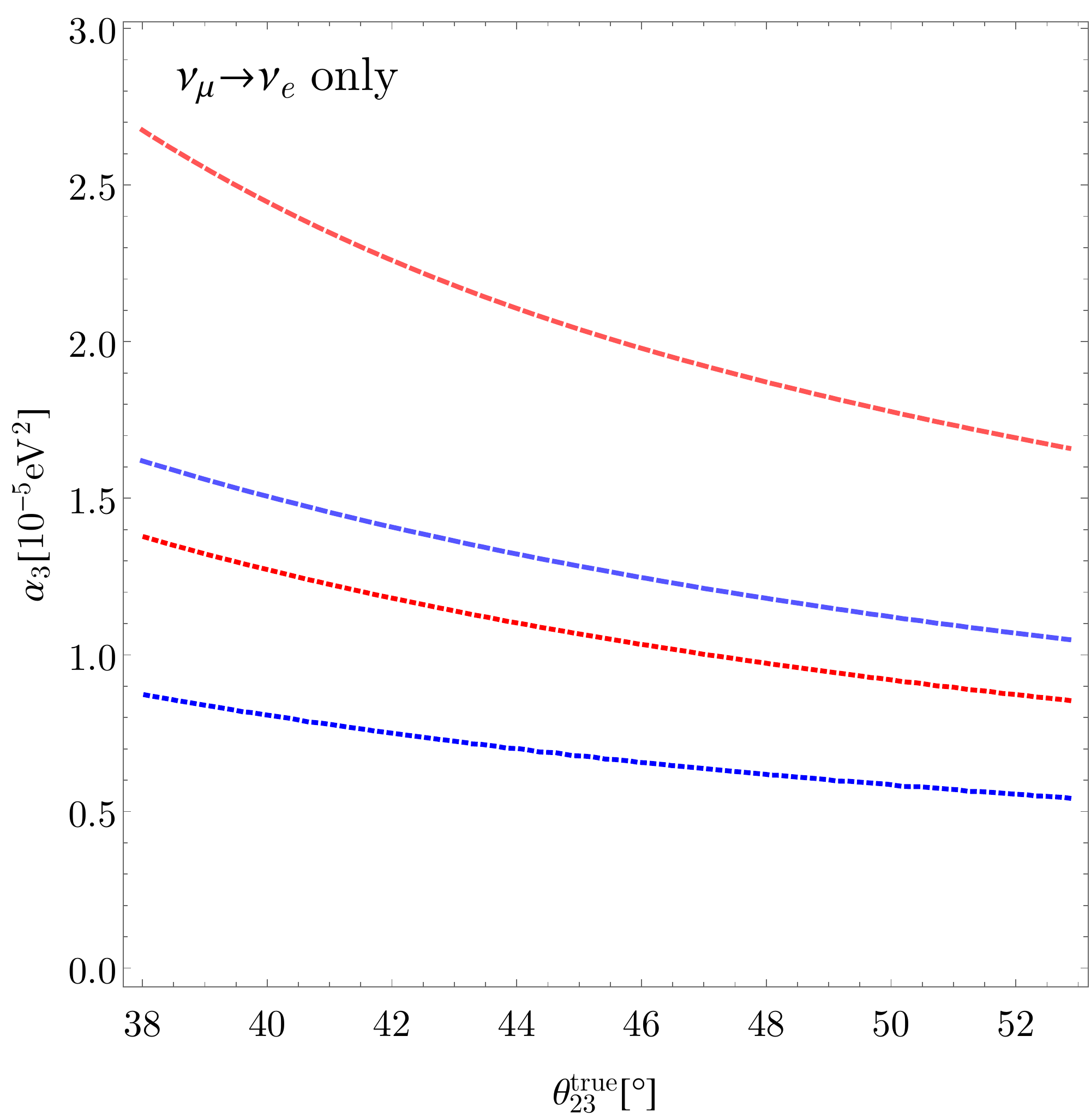} 
\caption{Sensitivity to $\alpha_3$ as a function of $\theta^{\rm true}_{23}$. Left: the sensitivity when combining $\nu_\mu$ disappearance and $\nu_e$ appearance, after both FHC and RHC runs. The curves show the sensitivity after marginalizing over $\delta^{\rm true}_{\rm CP}$, with the shaded area below giving the sensitivity for fixed values of the phase. Right: sensitivity curves for only FHC or RHC runs, including only the $\nu_e$ appearance channel.}
\label{fig:sens.th23}
\end{figure}
In Figure~\ref{fig:sens.th23}, we show the sensitivity to $\alpha_3$, for different values of a given $\theta_{23}^{\rm true}$. On the left panel, we use the full potential of DUNE by combining both $\nu_\mu$ disappearance and $\nu_e$ appearance, for both FHC and RHC modes. The sensitivity improves for larger values of $\sin^2\theta_{23}^{\rm true}$, since the VD contribution is proportional to this parameter. This means that for a fixed $\chi^2$, a larger $\theta^{\rm true}_{23}$ requires a smaller $\alpha_3$. We find that DUNE is sensitive at $3\sigma$ ($5\sigma$) to values of $\alpha_3$ around $(4-7)\times10^{-6}$~eV$^2$ ($(0.7-1.1)\times10^{-5}$~eV$^2$). Moreover, the shaded areas below the curves indicate the sensitivity for fixed values of the phase, that is, without marginalizing. We consider the full range of $\delta_{\rm CP}^{\rm true}$, and see that the uncertainty in this parameter has little impact on the sensitivity to $\alpha_3$.

The right panel of Figure~\ref{fig:sens.th23} shows the sensitivity using only the $\nu_e$ appearance channel, after marginalizing, for different modes. An analogous plot for the $\nu_\mu$ disappearance channel would have a much worse sensitivity to $\alpha_3$, which follows from our arguments in Section~\ref{sec:nevents}. As expected from our earlier discussion, both FHC and RHC curves have the same downward slope.

The $3\sigma$ ($5\sigma$) sensitivity for FHC is around $(1.0-1.6)\times10^{-5}$~eV$^2$ ($(1.6-2.7)\times10^{-5}$~eV$^2$). For RHC instead, the sensitivity is around $(5-9)\times10^{-6}$~eV$^2$ ($(0.8-1.4)\times10^{-5}$~eV$^2$). Thus, we see that sensitivity is determined principally by the RHC mode. This is easy to understand using the study of $(\Phi\times\sigma)$ in Section~\ref{sec:nevents}: the RHC flux is dominated by $\nu^{(+)}$, which have a smaller cross-section than $\nu^{(-)}$, such that the helicity-changing decays in VD have a greater impact.

\begin{figure}[tb]
\centering
\includegraphics[width=0.48\textwidth]{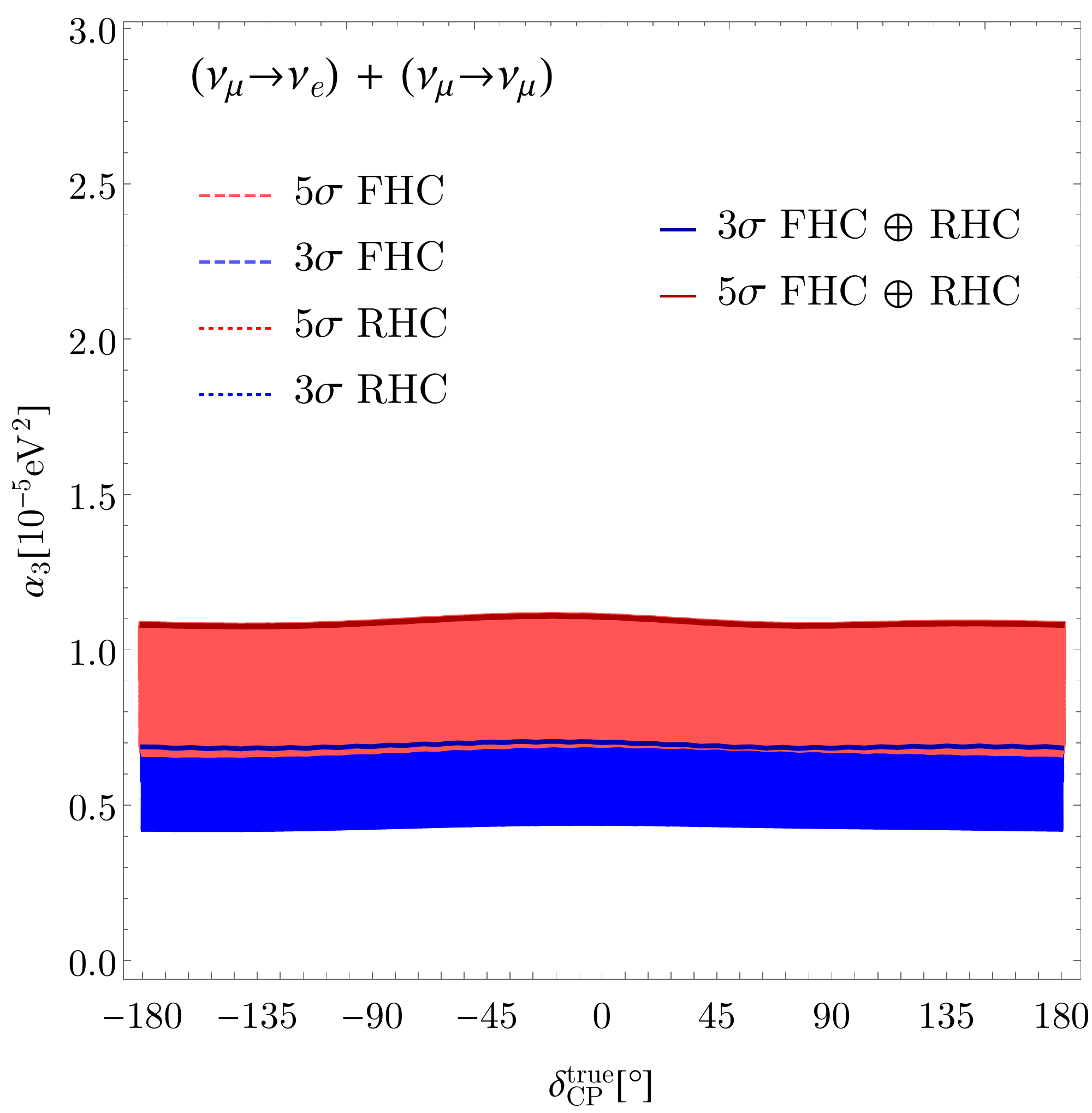} \hfill
\includegraphics[width=0.48\textwidth]{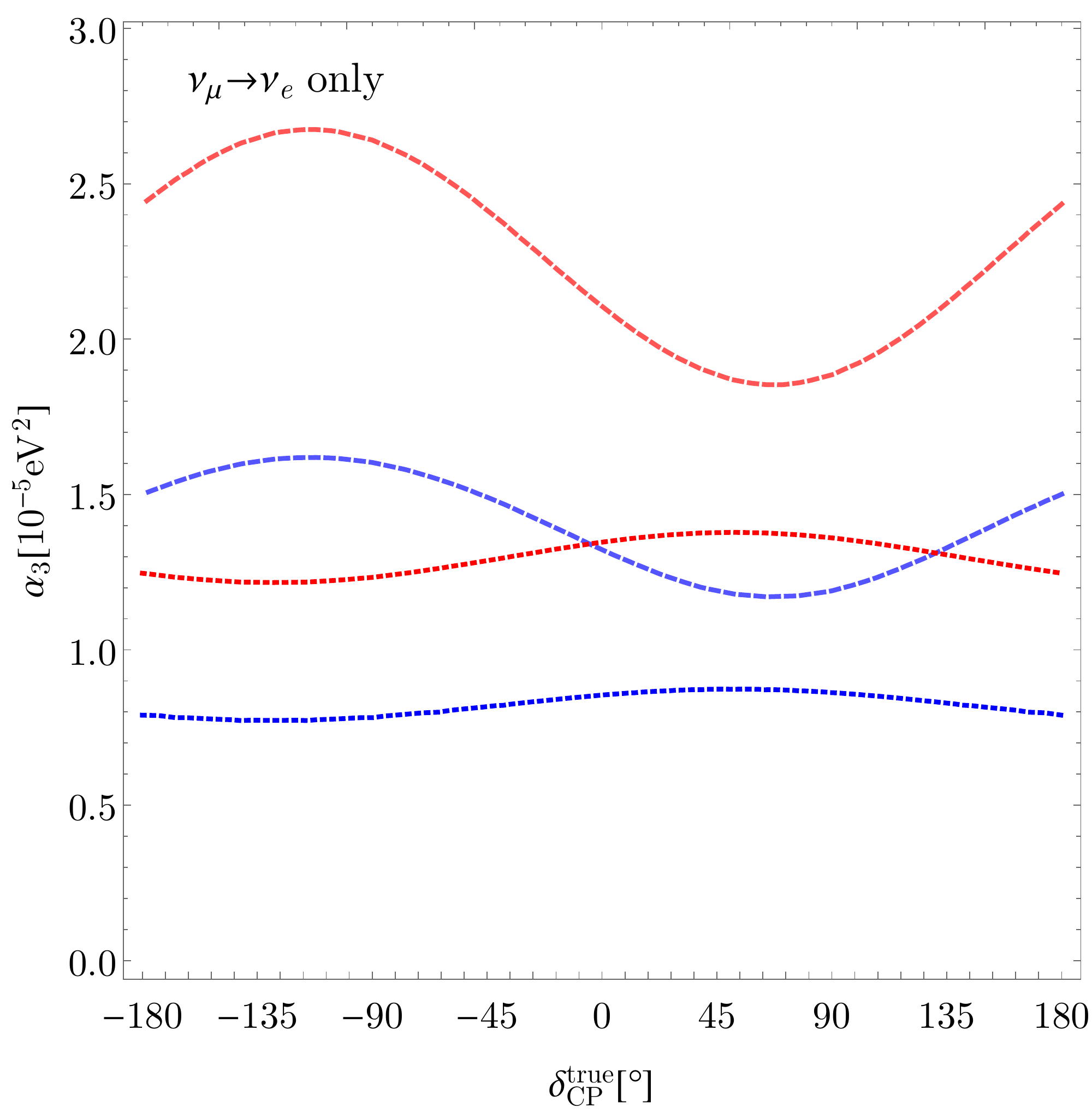}
\caption{As in Figure~\ref{fig:sens.th23}, but as a function of $\delta_{\rm CP}^{\rm true}$.}
\label{fig:sens.dCP}
\end{figure}
The sensitivity to $\alpha_3$ as a function of $\delta^{\rm true}_{\rm CP}$ is shown in Figure~\ref{fig:sens.dCP}. On the left panel, we have the same curves and shaded areas as in Figure~\ref{fig:sens.th23}, but this time marginalizing over $\theta_{23}^{\rm true}$. We find the curves to be almost flat, suggesting that the value of $\delta^{\rm true}_{\rm CP}$ is not important for the determination of $\alpha_3$.

Nevertheless, the right panel of Figure~\ref{fig:sens.dCP} shows a very different picture. Here, we find that the sensitivity has a very strong dependence on $\delta^{\rm true}_{\rm CP}$, particularly on the FHC mode. This, of course, is due to the influence of $\delta^{\rm true}_{\rm CP}$ on the total number of events. For example, for the FHC mode, a positive $\delta^{\rm true}_{\rm CP}$ would reduce the events coming from ID, which has a stronger dependence on $\delta^{\rm true}_{\rm CP}$ than VD. As a consequence of this reduction, a relatively small $\alpha_3$ is sufficient to generate a VD contribution that can be comparable to the ID component. This makes this point more sensitive to low values of $\alpha_3$. In contrast, a negative $\delta^{\rm true}_{\rm CP}$ implies a larger number of events expected from ID, such that a larger $\alpha_3$ is required to reach the same level of sensitivity compared to positive $\delta_{\rm CP}^{\rm true}$. Thus, the ratio between the ID and VD components has a relevant impact on the sensitivity.

The difference between the largest and smallest values of $\alpha_3$ in a sensitivity curve is most pronounced in the $5\sigma$ case of FHC, where the ID contribution clearly dominates (see Figure~\ref{fig:phisig.DUNE}), and is modulated by the value of $\delta_{\rm CP}^{\rm true}$. For the $3\sigma$ case the behaviour is the same, but the difference in $\alpha_3$ is diminished, due to the lesser number of ID events.

Due to CP conjugation, the situation for the RHC flux is opposite: scenarios with positive $\delta^{\rm true}_{\rm CP}$ are less sensitive to $\alpha_3$, in comparison to those with negative values. Here, the difference between the largest and smallest $\alpha_3$ is very small, that is, it has a milder dependence on $\delta_{\rm CP}^{\rm true}$, as the overall number of events is also small and the VD contribution is usually comparable to the ID one. When combining both FHC and RHC, we find that the latter has a stronger pull on the sensitivity. In addition, the curve averages out, leaving an almost flat result.

\begin{figure}[tb]
\centering
\includegraphics[width=0.48\textwidth]{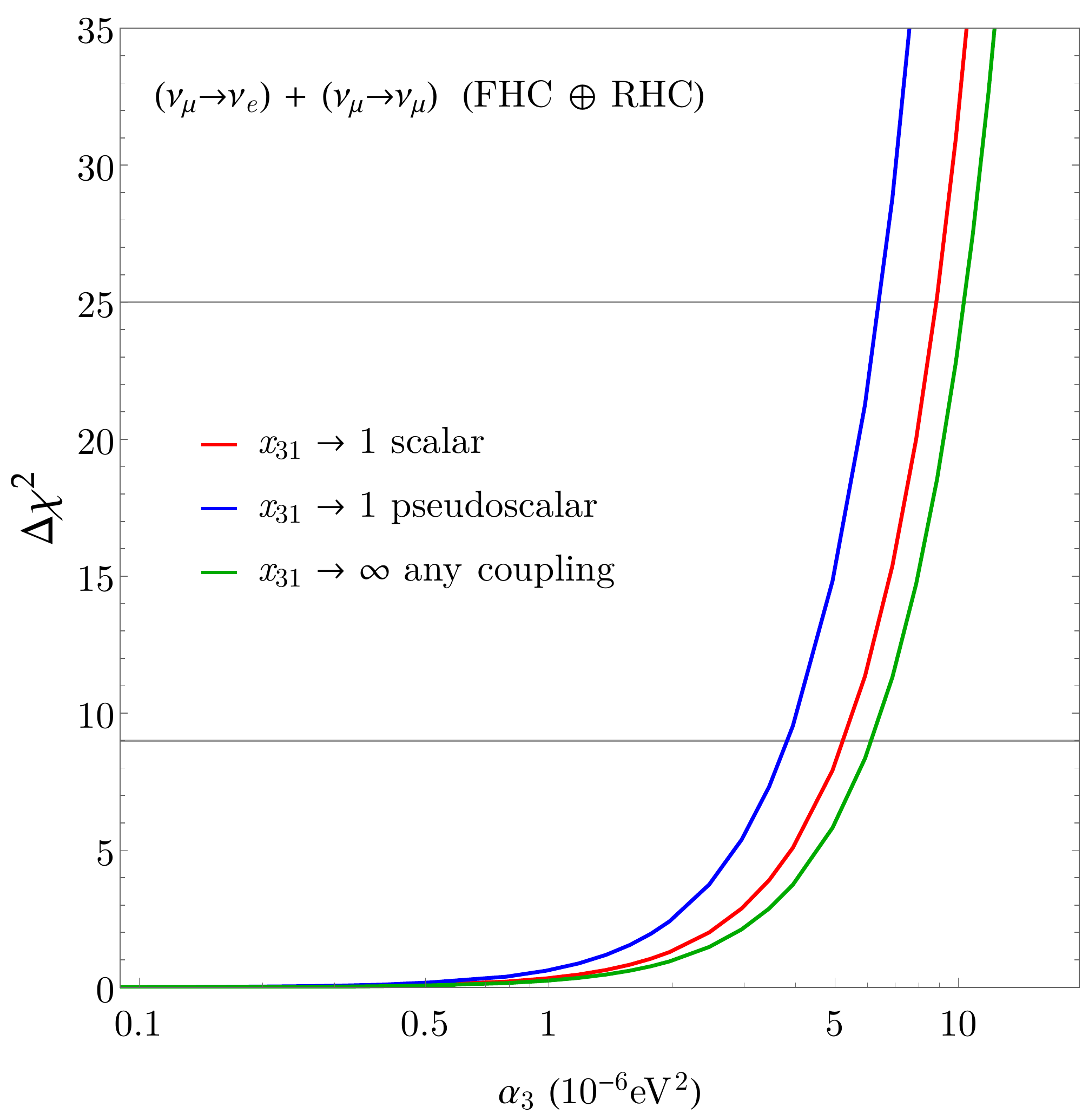}
\caption{Sensitivity to $\alpha_3$ in DUNE, combining $\nu_\mu$ disappearance and $\nu_e$ appearance, after both FHC and RHC runs. We marginalize over $\delta_{\rm CP}^{\rm true}$ and $\theta_{23}^{\rm true}$. The horizontal lines indicate the $3\sigma$ and $5\sigma$ confidence levels.}
\label{fig:sens.all}
\end{figure}

In Figure~\ref{fig:sens.all} we show the sensitivity to $\alpha_3$ for both $x_{31}\to1$ and $x_{31}\to\infty$ scenarios. As we have emphasized earlier, only in the former case it is necessary to distinguish between scalar and pseudoscalar couplings. In the Figure, the pseudoscalar coupling has a much better sensitivity, reaching values of $\alpha_3=3.8\times10^{-6}$~ eV$^2$ ($6.4\times10^{-6}$~eV$^2$) at $3\sigma\,(5\sigma)$, compared to the scalar coupling, which reaches $\alpha_3=5.2\times10^{-6}$~eV$^2$ ($8.8\times10^{-6}$~eV$^2$). This is due to the increased helicity-changing transitions that are typical of this coupling, increasing tensions with RHC expectations. In fact, for the pseudoscalar coupling, we find that the RHC-only curve clearly dominates the overall sensitivity. In contrast, the scalar coupling has less helicity-changing transitions, so the sensitivity does not improve as much. Nevertheless, the $x_{31}\to1$ scenario with scalar coupling has a better sensitivity than the $x_{31}\to\infty$ case, which reaches $\alpha_3=6.1\times10^{-6}$~eV$^2$ ($1.0\times10^{-5}$~eV$^2$). The reason for this is that, as we can see in Figure~\ref{fig:visible.compare}, the VD peak of the $x_{31}\to\infty$ scenario appears at very low values of energy, which are cut off by the experimental thresholds included in our simulation.

These numbers can be compared to other results in the literature, at $90\%$~C.L. For instance, for FD at T2K and MINOS~\cite{Gago:2017zzy}, the best limit is of $\alpha_3<5.6\times10^{-5}$~eV$^2$. The authors of~\cite{Choubey:2017dyu} work in the context of ID at DUNE, and obtain a sensitivity around $1.5\times10^{-5}$~eV$^2$. In contrast, the authors of~\cite{Coloma:2017zpg} take FD, and their best sensitivity is as low as $3.4\times10^{-6}$~eV$^2$. At this confidence level, our best sensitivity is of $2.0\times10^{-6}$~eV$^2$, which is comparable to the limit obtained with atmospheric neutrinos with ID, of $\alpha_3<2.2\times10^{-6}$~eV$^2$~\cite{GonzalezGarcia:2008ru}. 

\begin{figure}[tbp]
\centering
\includegraphics[width=0.49\textwidth]{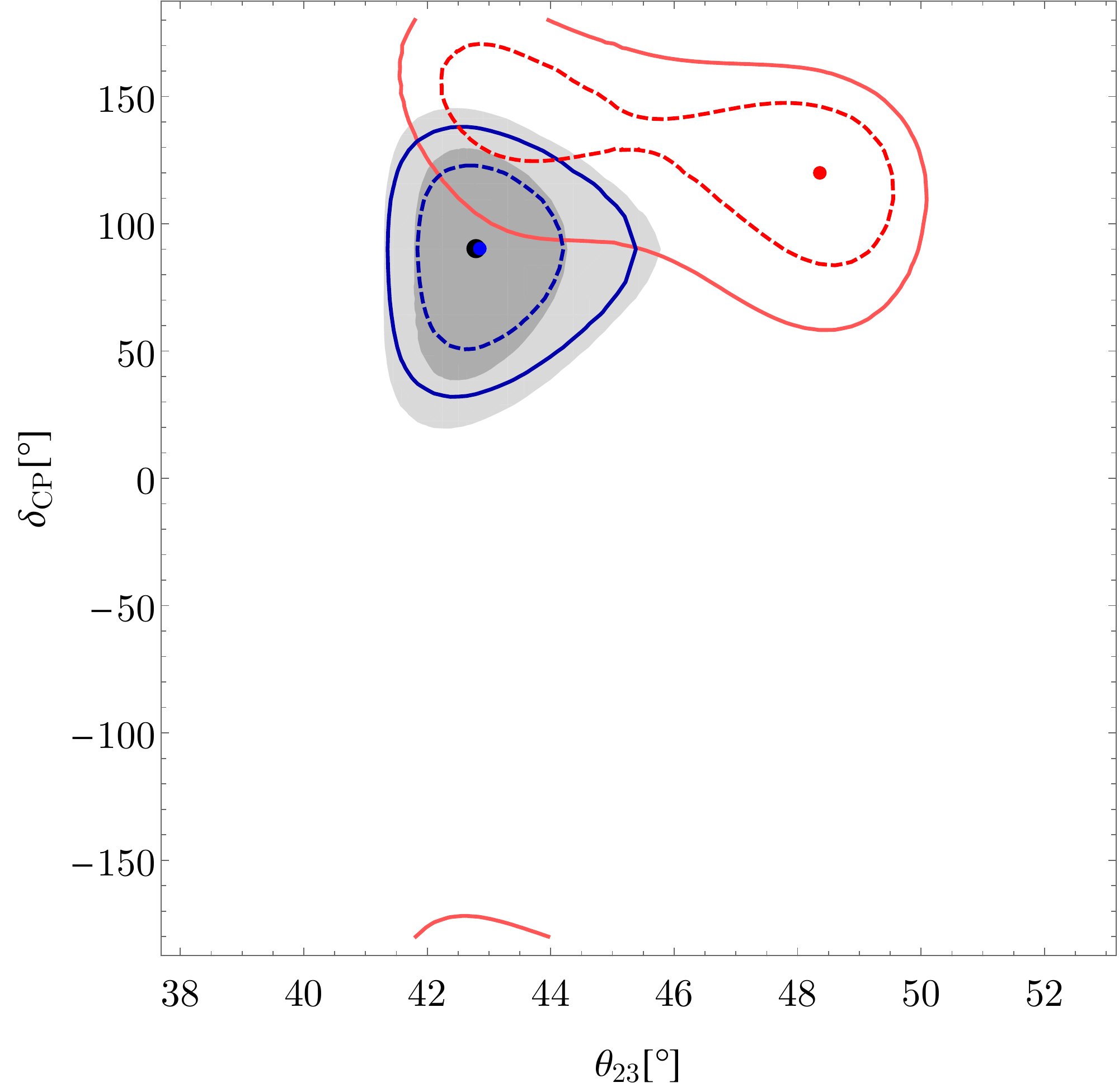} \hfill
\includegraphics[width=0.49\textwidth]{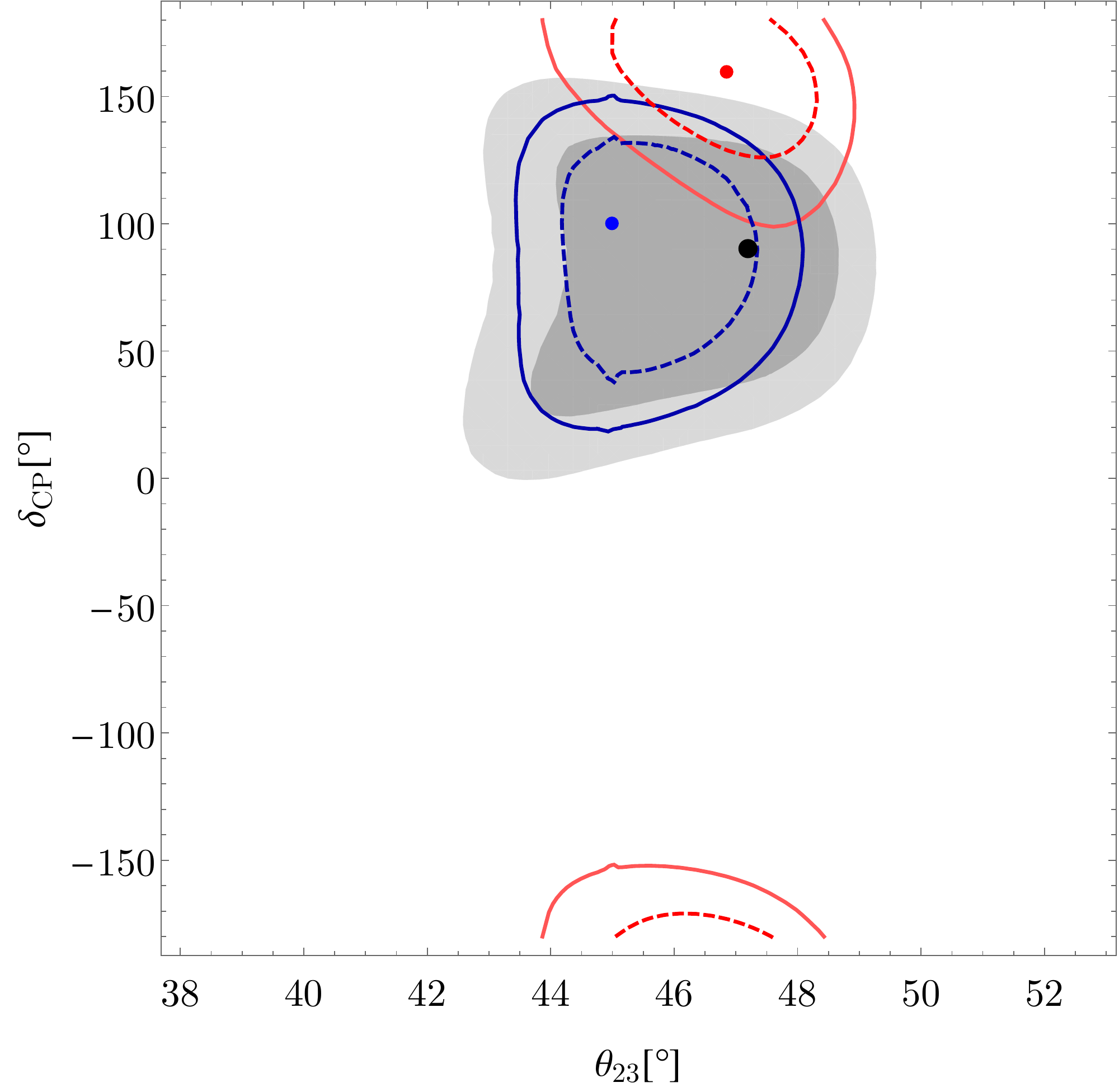} \\ 
\includegraphics[width=0.49\textwidth]{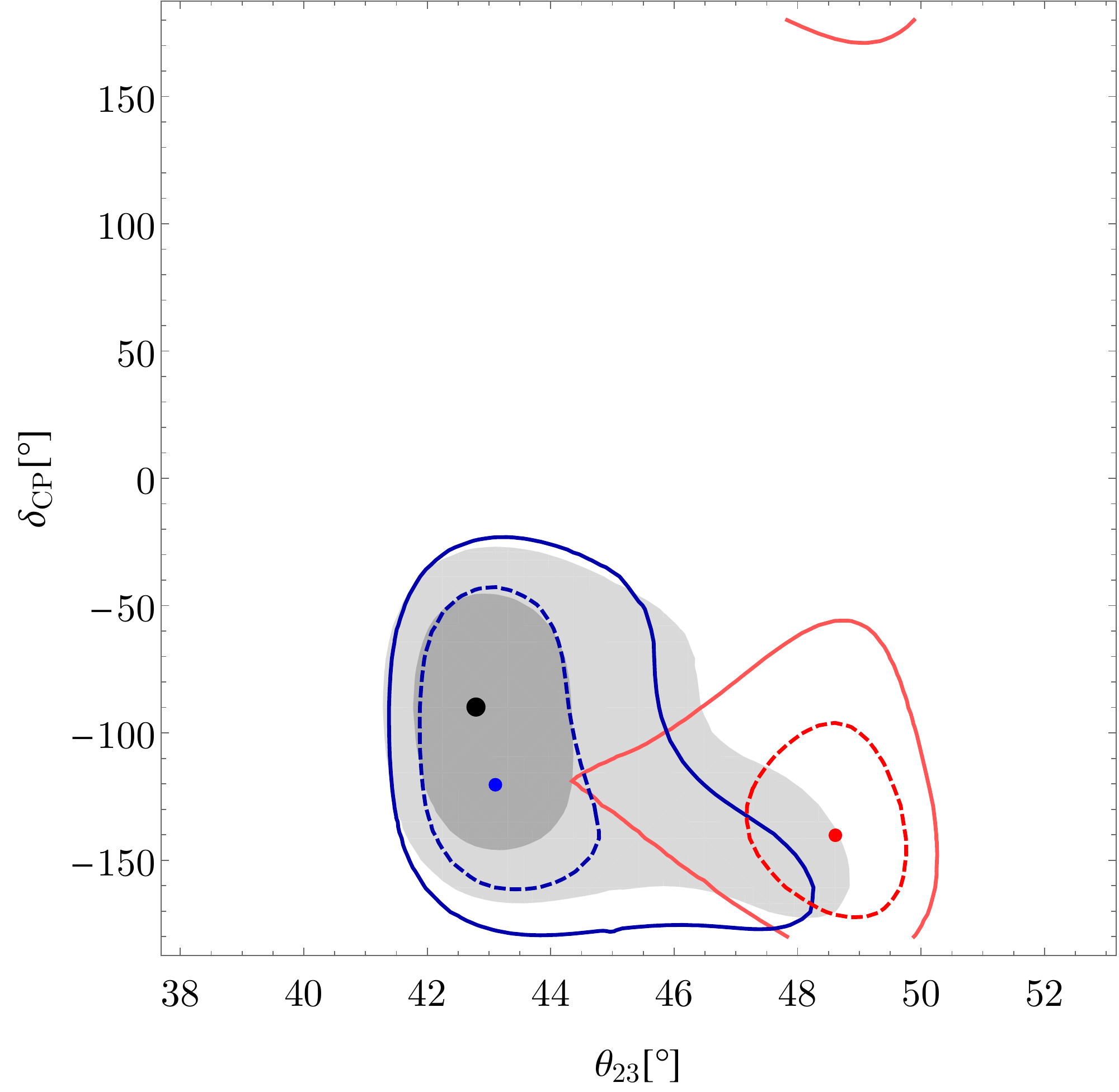} \hfill
\includegraphics[width=0.49\textwidth]{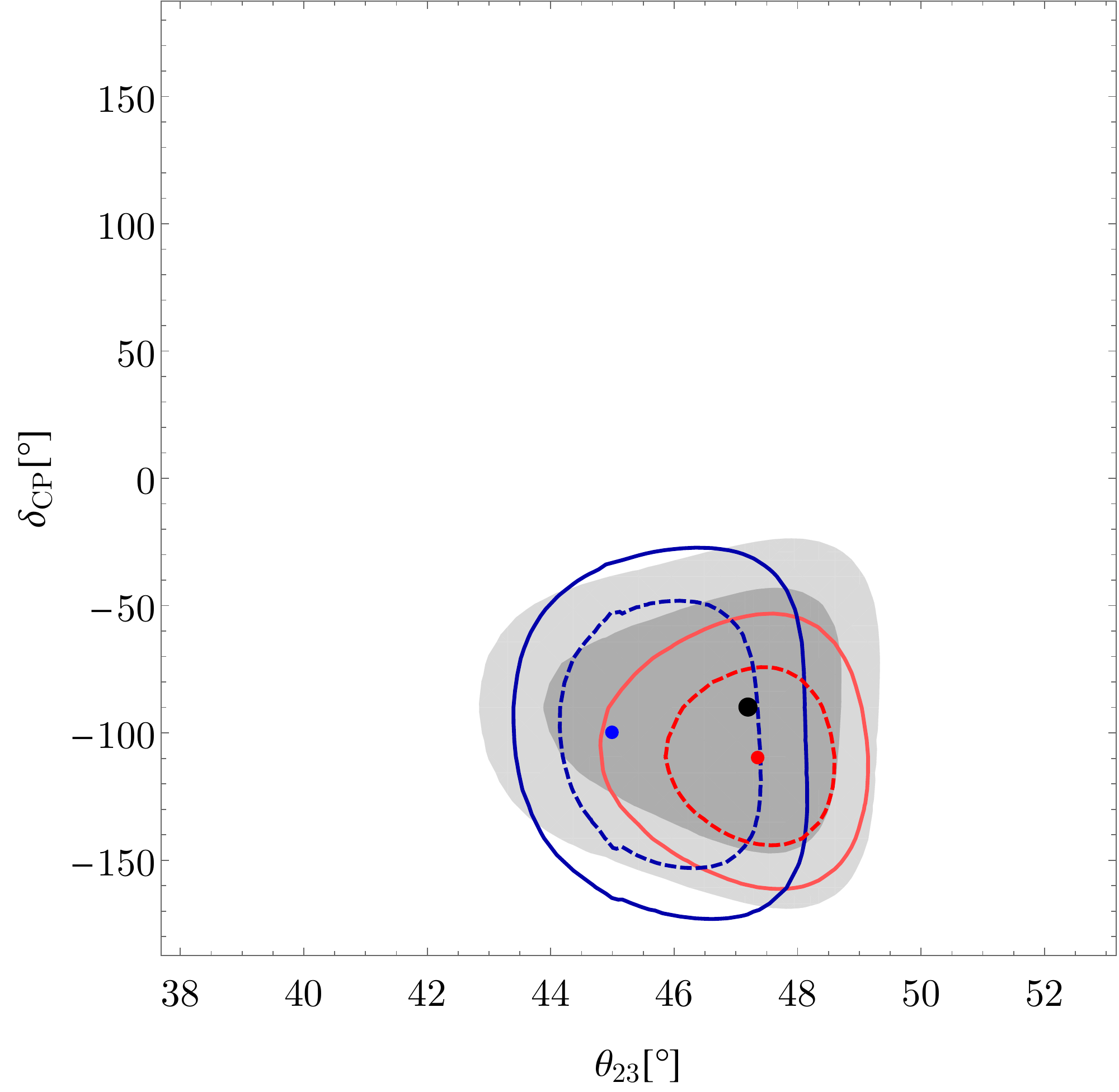}
\caption{Allowed regions, taking SO as a theoretical hypothesis, assuming that the measured signal was generated by SO (gray region), ID (blue curve) or FD (red curve). We include $\nu_\mu$ disappearance and $\nu_e$ appearance channels. The left (right) column takes $\theta_{23}^{\rm true}=42.8^\circ$ ($47.2^\circ$), while the upper (lower) row has $\delta^{\rm true}_{\rm CP}=90^{\circ}$ ($-90^\circ$). We take $\alpha^{\rm true}_3=4\times10^{-5}$~eV$^2$. Dashed and solid lines (dark and light regions) correspond to $3\sigma$ and $5\sigma$ confidence levels, respectively, with the dots indicating the best-fit points.}
\label{fig:fit}
\end{figure}

For the final part of our analysis, we compare the impact of the SO, ID and FD scenarios in the determination of $\theta_{23}$ and $\delta_{\rm CP}$. For each scenario we generate a specific number of events by fixing oscillation and decay parameters in combinations of {\it true} values: $\theta_{23}^{\rm true}=\{42.8^\circ,\,47.2^\circ\}$, $\delta^{\rm true}_{\rm CP}=\{90^{\circ},\,-90^\circ\}$. For the ID and FD scenarios, we set $\alpha^{\rm true}_3=4\times10^{-5}$~eV$^2$, as was done on Section~\ref{sec:nevents}. When performing the fit, we assume SO as a theoretical hypothesis, that is, we evaluate:
\begin{equation}
\chi^2(\theta_{23},\,\delta_{\rm CP},\,0,\,\theta^{\rm true}_{23},\,\delta^{\rm true}_{\rm CP},\,\alpha_3^{\rm true})
\end{equation}
where, of course, $\alpha_3^{\rm true}=0$~eV$^2$ in the SO scenario. We include both $\nu_\mu$ disappearance and $\nu_e$ appearance, with both FHC and RHC runs.

Results are shown in Figure~\ref{fig:fit}. We find small differences between SO and ID, which is due to $\alpha_3$ not being large enough to have any effect. The reason for including these curves is mainly for comparison with FD, which is the main focus of this work. If we had used a larger $\alpha_3$, we would have found a stronger impact on the determination of $\theta_{23}$, as was reported in detail in~\cite{Choubey:2017dyu}.

On the other hand, the FD scenario is much more susceptible to values of $\alpha_3$ of this order of magnitude. This is consistent with Figure~\ref{fig:phisig.DUNE}, that is, that the additional VD component can dominate the flux at low energy, which in turn is the main reason for the increased sensitivity shown in Figure~\ref{fig:sens.all}.

The regions can be understood by comparing the impact of FD on $\nu_\mu$ disappearance and $\nu_e$ appearance. First, since we know from Section~\ref{sec:nevents} that $\nu_\mu$ disappearance is not strongly affected by FD, we find the same kind of constraints on the minimum and maximum possible values of $\theta_{23}$, including the octant degeneracy~\cite{Barger:2001yr,BurguetCastell:2002qx,Gago:2006rb}.

The important modification on the fit happen because of the $\nu_e$ appearance channel. As reported in Section~\ref{sec:nevents}, the inclusion of VD leads to an increase in events in both FHC and RHC modes within our simulated data sample. This fact is the reason why the SO fit in general prefers larger values of $\theta_{23}$, since the leading term in the $\nu_\mu\to\nu_e$ oscillation probability is proportional to $\sin^2\theta_{23}$. In fact, {\it true} points generated with FD in the lower octant of $\theta_{23}$ can have an SO solution on the higher octant. For {\it true} points generated in the higher octant, the $\nu_\mu$ disappearance constraint does not allow $\theta_{23}$ to increase.

Our {\it true} points were generated for values of $\delta^{\rm true}_{\rm CP}$ where the CP asymmetry is maximal. However, in the fit the SO regions tend to prefer values of $\delta_{\rm CP}$ closer to $\pm\pi$, such that the asymmetry is diminished.

\section{Conclusions}

Starting from the treatment given in~\cite{Lindner:2001fx}, we have formulated a description for matter effects in visible neutrino decay. In order to understand these effects, we have implemented a $(\Phi\times\sigma)$ study in two scenarios. On the first one, we use the flux and baseline for DUNE, while on the second one we use the same flux, but consider the corresponding baseline for ANDES. For DUNE, we find that only the ID component of $\nu_e$ appearance can be affected by matter, the effect for all other components, such as VD at $\nu_e$ appearance, can be completely ignored. In contrast, for ANDES, not only do we have an enhanced effect on the ID component due to matter, but also find that the VD component receives a relevant modification. This is especially important for $\nu_e$ appearance, but the effects can also be seen on $\nu_\mu$ disappearance.

Another important part of this work was devoted to the calculation of the sensitivity of DUNE to $\alpha_3$. To this end, we performed as very detailed simulation of DUNE, described in Section~\ref{sec:paramfits}, using the publicly available files of the Collaboration~\cite{Alion:2016uaj}. On that section, we showed the dependence of the sensitivity on $\theta_{23}$ and $\delta_{\rm CP}$, using both $\nu_\mu$ disappearance and $\nu_e$ appearance channels, and both FHC and RHC modes. Our final sensitivity to $\alpha_3$ depended on the lightest neutrino mass (encoded on the value of $x_{31}$) and on the type of coupling between the neutrino and the Majoron (scalar or pseudoscalar). For the $x_{31}\to1$ scenario, we found the sensitivity of DUNE to be:
\begin{align}
 \alpha^{(s)}_3 < 2.8\times10^{-6}{~\rm eV}^2~, & & \alpha^{(p)}_3 < 2.0\times10^{-6}{~\rm eV}^2~
\end{align}
while for $x_{31}\to\infty$:
\begin{equation}
 \alpha_3^{(s,\,p)}<3.2\times10^{-6}{~\rm eV}^2
\end{equation}
We note that these values of sensitivity are the best ones obtained so far for long-baseline experiments, and are comparable to the current limits set by atmospheric neutrinos using ID~\cite{GonzalezGarcia:2008ru}.

Finally, in order to understand the impact of a non-zero $\alpha_3$ on the determination of oscillation parameters, we performed a fit on $\theta_{23}$ and $\delta_{\rm CP}$ assuming SO, with data generated for the FD scenario. We found that the allowed regions would shift towards larger values of $\theta_{23}$, and towards CP-conserving values of $\delta_{\rm CP}$.

\section{Acknowledgements}

A.M.G.~and J.J.P.~acknowledge funding by the {\it Direcci\'on de Gesti\'on de la Investigaci\'on} at PUCP, through grant DGI-2015-3-0026. M.V.A.S.~acknowledges funding by CienciActiva-CONCYTEC Grants 026-2015 and CONV-236-2015-FONDECYT. A.M.C.C.~received funding from CienciActiva-CONCYTEC Grant 233-2015-1. We would like to thank F.~D\'iaz-Desposorio for his help with the GLoBES implementation, and O.~A.~D\'iaz in the {\it Direcci\'on de Tecnolog\'ias de Informaci\'on} for support within the {\tt LEGION} system. 

\bibliographystyle{epjc}
\bibliography{nudecay} 

\begin{thebibliography}{10}
\providecommand{\url}[1]{\texttt{#1}}
\providecommand{\urlprefix}{URL }
\providecommand{\eprint}[2][]{\url{#2}}

\bibitem{Gago:2000qc}
A.~M. Gago, E.~M. Santos, W.~J.~C. Teves, et~al., Phys. Rev. \textbf{D63},
  073001 (2001), \eprint{arXiv:hep-ph/0009222}

\bibitem{Gago:2002na}
A.~M. Gago, E.~M. Santos, W.~J.~C. Teves, et~al.  (2002),
  \eprint{arXiv:hep-ph/0208166}

\bibitem{Fogli:2003th}
G.~Fogli, E.~Lisi, A.~Marrone, et~al., Phys. Rev. \textbf{D67}, 093006 (2003),
  \eprint{arXiv:hep-ph/0303064}

\bibitem{Morgan:2004vv}
D.~Morgan, E.~Winstanley, J.~Brunner, et~al., Astropart. Phys. \textbf{25}, 311
  (2006), \eprint{arXiv:astro-ph/0412618}

\bibitem{Lisi:2000zt}
E.~Lisi, A.~Marrone, D.~Montanino, Phys. Rev. Lett. \textbf{85}, 1166 (2000),
  \eprint{arXiv:hep-ph/0002053}

\bibitem{Hooper:2004xr}
D.~Hooper, D.~Morgan, E.~Winstanley, Phys. Lett. \textbf{B609}, 206 (2005),
  \eprint{hep-ph/0410094}

\bibitem{Farzan:2008zv}
Y.~Farzan, T.~Schwetz, A.~Y. Smirnov, JHEP \textbf{0807}, 067 (2008),
  \eprint{arXiv:0805.2098}

\bibitem{Oliveira:2010zzd}
R.~L.~N. Oliveira, M.~M. Guzzo, Eur.Phys.J. \textbf{C69}, 493 (2010)

\bibitem{robertoPhd}
R.~L.~N. Oliveira, \emph{{Dissipa\c{c}\~ao qu\^antica em oscila\c{c}\~oes de
  neutrinos}}, Ph.D. thesis, UNICAMP (2013)

\bibitem{Oliveira:2013nua}
R.~L.~N. Oliveira, M.~M. Guzzo, Eur.Phys.J. \textbf{C73}, 2434 (2013)

\bibitem{Berryman:2014yoa}
J.~M. Berryman, A.~de~Gouv\^ea, D.~Hern\'andez, et~al., Phys. Lett.
  \textbf{B742}, 74 (2015), \eprint{arXiv:1407.6631}

\bibitem{Carpio:2017nui}
J.~A. Carpio, E.~Massoni, A.~M. Gago  (2017), \eprint{1711.03680}

\bibitem{Coloma:2018idr}
P.~Coloma, J.~Lopez-Pavon, I.~Martinez-Soler, et~al.  (2018),
  \eprint{1803.04438}

\bibitem{GonzalezGarcia:1998hj}
M.~C. Gonzalez-Garcia, M.~M. Guzzo, P.~Krastev, et~al., Phys.Rev.Lett.
  \textbf{82}, 3202 (1999), \eprint{arXiv:hep-ph/9809531}

\bibitem{Bergmann:2000gp}
S.~Bergmann, M.~M. Guzzo, P.~C. de~Holanda, et~al., Phys.Rev. \textbf{D62},
  073001 (2000)

\bibitem{Guzzo:2004ue}
M.~M. Guzzo, P.~C. de~Holanda, O.~L.~G. Peres, Phys.Lett. \textbf{B591}, 1
  (2004), \eprint{arXiv:hep-ph/0403134}

\bibitem{Gago:2001si}
A.~M. Gago, M.~M. Guzzo, P.~C. de~Holanda, et~al., Phys. Rev. \textbf{D65},
  073012 (2002), \eprint{arXiv:hep-ph/0112060}

\bibitem{Gago:2001xg}
A.~M. Gago, M.~M. Guzzo, H.~Nunokawa, et~al., Phys. Rev. \textbf{D64}, 073003
  (2001), \eprint{arXiv:hep-ph/0105196}

\bibitem{Fogli:2007tx}
G.~L. Fogli, E.~Lisi, A.~Marrone, et~al., Phys. Rev. \textbf{D76}, 033006
  (2007), \eprint{arXiv:0704.2568}

\bibitem{Ohlsson:2012kf}
T.~Ohlsson, Rept. Prog. Phys. \textbf{76}, 044201 (2013),
  \eprint{arXiv:1209.2710}

\bibitem{Esmaili:2013fva}
A.~Esmaili, A.~Y. Smirnov, JHEP \textbf{1306}, 026 (2013),
  \eprint{arXiv:1304.1042}

\bibitem{Frieman:1987as}
J.~A. Frieman, H.~E. Haber, K.~Freese, Phys.Lett. \textbf{B200}, 115 (1988)

\bibitem{Raghavan:1987uh}
R.~Raghavan, X.-G. He, S.~Pakvasa, Phys.Rev. \textbf{D38}, 1317 (1988)

\bibitem{Berezhiani:1991vk}
Z.~Berezhiani, G.~Fiorentini, M.~Moretti, et~al., Z.Phys. \textbf{C54}, 581
  (1992)

\bibitem{Berezhiani:1992ry}
Z.~Berezhiani, G.~Fiorentini, A.~Rossi, et~al., JETP Lett. \textbf{55}, 151
  (1992)

\bibitem{Berezhiani:1993iy}
Z.~G. Berezhiani, A.~Rossi, in \emph{{5th International Symposium on Neutrino
  Telescopes Venice, Italy, March 2-4, 1993}}, 123--135 (1993), [,123(1993)],
  \eprint{arXiv:hep-ph/9306278}

\bibitem{Barger:1999bg}
V.~D. Barger, J.~Learned, P.~Lipari, et~al., Phys.Lett. \textbf{B462}, 109
  (1999), \eprint{arXiv:hep-ph/9907421}

\bibitem{Lindner:2001fx}
M.~Lindner, T.~Ohlsson, W.~Winter, Nucl. Phys. \textbf{B607}, 326 (2001),
  \eprint{arXiv:hep-ph/0103170}

\bibitem{Beacom:2002cb}
J.~F. Beacom, N.~F. Bell, Phys.Rev. \textbf{D65}, 113009 (2002),
  \eprint{arXiv:hep-ph/0204111}

\bibitem{Joshipura:2002fb}
A.~S. Joshipura, E.~Masso, S.~Mohanty, Phys.Rev. \textbf{D66}, 113008 (2002),
  \eprint{arXiv:hep-ph/0203181}

\bibitem{Bandyopadhyay:2002qg}
A.~Bandyopadhyay, S.~Choubey, S.~Goswami, Phys.Lett. \textbf{B555}, 33 (2003),
  \eprint{arXiv:hep-ph/0204173}

\bibitem{Ando:2004qe}
S.~Ando, Phys.Rev. \textbf{D70}, 033004 (2004), \eprint{arXiv:hep-ph/0405200}

\bibitem{Fogli:2004gy}
G.~Fogli, E.~Lisi, A.~Mirizzi, et~al., Phys.Rev. \textbf{D70}, 013001 (2004),
  \eprint{arXiv:hep-ph/0401227}

\bibitem{PalomaresRuiz:2005vf}
S.~Palomares-Ruiz, S.~Pascoli, T.~Schwetz, JHEP \textbf{0509}, 048 (2005),
  \eprint{arXiv:hep-ph/0505216}

\bibitem{GonzalezGarcia:2008ru}
M.~C. Gonzalez-Garcia, M.~Maltoni, Phys. Lett. \textbf{B663}, 405 (2008),
  \eprint{arXiv:0802.3699}

\bibitem{Maltoni:2008jr}
M.~Maltoni, W.~Winter, JHEP \textbf{0807}, 064 (2008), \eprint{arXiv:0803.2050}

\bibitem{Baerwald:2012kc}
P.~Baerwald, M.~Bustamante, W.~Winter, JCAP \textbf{1210}, 020 (2012),
  \eprint{arXiv:1208.4600}

\bibitem{Meloni:2006gv}
D.~Meloni, T.~Ohlsson, Phys.Rev. \textbf{D75}, 125017 (2007),
  \eprint{arXiv:hep-ph/0612279}

\bibitem{Das:2010sd}
C.~R. Das, J.~Pulido, Phys.Rev. \textbf{D83}, 053009 (2011)

\bibitem{Dorame:2013lka}
L.~Dorame, O.~G. Miranda, J.~W.~F. Valle, Front.in Phys. \textbf{1}, 25 (2013),
  \eprint{arXiv:1303.4891}

\bibitem{Gomes:2014yua}
R.~A. Gomes, A.~L.~G. Gomes, O.~L.~G. Peres, Phys. Lett. \textbf{B740}, 345
  (2015), \eprint{arXiv:1407.5640}

\bibitem{Berryman:2014qha}
J.~M. Berryman, A.~de~Gouv\^ea, D.~Hernandez, Phys. Rev. \textbf{D92}, 7,
  073003 (2015), \eprint{arXiv:1411.0308}

\bibitem{Picoreti:2015ika}
R.~Picoreti, M.~M. Guzzo, P.~C. de~Holanda, et~al., Phys. Lett. \textbf{B761},
  70 (2016), \eprint{arXiv:1506.08158}

\bibitem{Abrahao:2015rba}
T.~Abrah\~ao, H.~Minakata, H.~Nunokawa, et~al., JHEP \textbf{11}, 001 (2015),
  \eprint{arXiv:1506.02314}

\bibitem{Bustamante:2016ciw}
M.~Bustamante, J.~F. Beacom, K.~Murase, Phys. Rev. \textbf{D95}, 6, 063013
  (2017), \eprint{arXiv:1610.02096}

\bibitem{Gago:2017zzy}
A.~M. Gago, R.~A. Gomes, A.~L.~G. Gomes, et~al., JHEP \textbf{11}, 022 (2017),
  \eprint{1705.03074}

\bibitem{Coloma:2017zpg}
P.~Coloma, O.~L.~G. Peres  (2017), \eprint{1705.03599}

\bibitem{Choubey:2017dyu}
S.~Choubey, S.~Goswami, D.~Pramanik, JHEP \textbf{02}, 055 (2018),
  \eprint{1705.05820}

\bibitem{Choubey:2017eyg}
S.~Choubey, S.~Goswami, C.~Gupta, et~al., Phys. Rev. \textbf{D97}, 3, 033005
  (2018), \eprint{1709.10376}

\bibitem{Kachelriess:2000qc}
M.~Kachelriess, R.~Tomas, J.~W.~F. Valle, Phys. Rev. \textbf{D62}, 023004
  (2000), \eprint{arXiv:hep-ph/0001039}

\bibitem{Pasquini:2015fjv}
P.~S. Pasquini, O.~L.~G. Peres, Phys. Rev. \textbf{D93}, 5, 053007 (2016),
  [Erratum: Phys. Rev.D93,no.7,079902(2016)], \eprint{arXiv:1511.01811}

\bibitem{Gando:2012pj}
A.~Gando, et~al. (KamLAND-Zen), Phys. Rev. \textbf{C86}, 021601 (2012),
  \eprint{arXiv:1205.6372}

\bibitem{Agostini:2015nwa}
M.~Agostini, et~al., Eur. Phys. J. \textbf{C75}, 9, 416 (2015),
  \eprint{arXiv:1501.02345}

\bibitem{Hannestad:2005ex}
S.~Hannestad, G.~Raffelt, Phys. Rev. \textbf{D72}, 103514 (2005),
  \eprint{arXiv:hep-ph/0509278}

\bibitem{Archidiacono:2013dua}
M.~Archidiacono, S.~Hannestad, JCAP \textbf{1407}, 046 (2014),
  \eprint{arXiv:1311.3873}

\bibitem{Moss:2017pur}
Z.~Moss, M.~H. Moulai, C.~A. Argüelles, et~al., Phys. Rev. \textbf{D97}, 5,
  055017 (2018), \eprint{1711.05921}

\bibitem{Acciarri:2015uup}
R.~Acciarri, et~al. (DUNE)  (2015), \eprint{1512.06148}

\bibitem{Bertou:2012fk}
X.~Bertou, Eur. Phys. J. Plus \textbf{127}, 104 (2012)

\bibitem{Wolfenstein:1977ue}
L.~Wolfenstein, Phys. Rev. \textbf{D17}, 2369 (1978), [,294(1977)]

\bibitem{Mikheev:1986gs}
S.~P. Mikheev, A.~{\relax Yu}. Smirnov, Sov. J. Nucl. Phys. \textbf{42}, 913
  (1985), [,305(1986)]

\bibitem{Ade:2013zuv}
P.~A.~R. Ade, et~al. (Planck), Astron. Astrophys. \textbf{571}, A16 (2014),
  \eprint{arXiv:1303.5076}

\bibitem{Alion:2016uaj}
T.~Alion, et~al. (DUNE)  (2016), \eprint{1606.09550}

\bibitem{Strait:2016mof}
J.~Strait, et~al. (DUNE)  (2016), \eprint{1601.05823}

\bibitem{Acciarri:2016ooe}
R.~Acciarri, et~al. (DUNE)  (2016), \eprint{1601.02984}

\bibitem{Machado:2012ee}
P.~A.~N. Machado, T.~Muhlbeier, H.~Nunokawa, et~al., Phys. Rev. \textbf{D86},
  125001 (2012), \eprint{1207.5454}

\bibitem{Last}
X.~Bertou, Conference APS April Meeting  (2017),
  \urlprefix\url{http://andeslab.org/pdf/ANDES-Talk2017.pdf}

\bibitem{andespage}
 \urlprefix\url{http://www.andeslab.org/index.php?lang=uk}

\bibitem{Dziewonski:1981xy}
A.~M. Dziewonski, D.~L. Anderson, Phys. Earth Planet. Interiors \textbf{25},
  297 (1981)

\bibitem{Esteban:2016qun}
I.~Esteban, M.~C. Gonzalez-Garcia, M.~Maltoni, et~al., JHEP \textbf{01}, 087
  (2017), \eprint{1611.01514}

\bibitem{Huber:2004ka}
P.~Huber, M.~Lindner, W.~Winter, Comput. Phys. Commun. \textbf{167}, 195
  (2005), \eprint{hep-ph/0407333}

\bibitem{Huber:2007ji}
P.~Huber, J.~Kopp, M.~Lindner, et~al., Comput. Phys. Commun. \textbf{177}, 432
  (2007), \eprint{hep-ph/0701187}

\bibitem{An:2012eh}
F.~P. An, et~al. (Daya Bay), Phys. Rev. Lett. \textbf{108}, 171803 (2012),
  \eprint{arXiv:1203.1669}

\bibitem{Barger:2001yr}
V.~Barger, D.~Marfatia, K.~Whisnant, Phys. Rev. \textbf{D65}, 073023 (2002),
  \eprint{hep-ph/0112119}

\bibitem{BurguetCastell:2002qx}
J.~Burguet-Castell, M.~B. Gavela, J.~J. Gomez-Cadenas, et~al., Nucl. Phys.
  \textbf{B646}, 301 (2002), \eprint{hep-ph/0207080}

\bibitem{Gago:2006rb}
A.~M. Gago, J.~Jones-Perez, Phys. Rev. \textbf{D75}, 033004 (2007),
  \eprint{hep-ph/0611110}

\end{thebibliography}
 
\end{document}